\documentclass[twocolumn,prb,showpacs,aps,floats,amsmath,amssymb]{revtex4}

\usepackage{color}
\usepackage{graphicx}
\usepackage{dcolumn}
\usepackage{bm}
\usepackage{mciteplus}
\usepackage{float}

\begin{document}

\title{Spin-glass behavior in Ni-doped
La$_{1.85}$Sr$_{0.15}$CuO$_{4}$}

\author{A. Malinowski}
\author{V. L. Bezusyy}
\author{R. Minikayev}
\author{P. Dziawa}
\author{Y. Syryanyy}
\author{ M. Sawicki}
\affiliation{Institute of Physics, Polish Academy of Sciences,
02-668 Warsaw, Poland}
\date{\today}

\begin{abstract}
The dynamic and static magnetic properties of
La$_{1.85}$Sr$_{0.15}$Cu$_{1-y}$Ni$_{y}$O$_{4}$ with Ni
concentration up to $y$=0.63 are reported. All the features that
characterize the spin-glass (SG) behavior are found: bifurcation
of the dc susceptibility $\chi$ vs temperature curve, a peak in
the zero-field cooling branch of this curve accompanied by a step
in the imaginary part of the ac susceptibility $\chi_{ac}$,
frequency dependence of the peak position in the real part of
$\chi_{ac}$ and scaling behavior. The decay of remnant
magnetization is described by a stretched-exponential function.
The characteristic time from the critical slowing-down formula
that governs the dynamics of the system suggests the existence of
the spin clusters. The strongest interactions between the
fluctuating entities are at $y$ equal to the charge carriers
concentration in the system. The SG transition temperature
decreases linearly with decreasing $y$ for $y$$\leq$0.30 and
extrapolates to 0 K at $y$=0 what means that Ni display a magnetic
character in the surrounding Cu-O network starting from the
smallest concentration $y$. The static critical exponents
characterizing the scaling behavior of the nonlinear part of
$\chi$ lie between those typical for three dimensional Ising-like
and Heisenberg-like systems.
\end{abstract}

\maketitle

\section{Introduction}

Transition-metal oxides (TMO) seem to be a nature's playground
where a strong interplay between spin, charge, and lattice degrees
of freedom appears. When doped with holes, their phase diagram
involves spin- and charge-density-waves, Jahn-Teller distortions,
ferroelectricity and superconductivity, to mention only few of
their complex ground states.\cite{Bishop2004} The spin-glass (SG)
phase is also quite common to the phase diagram of TMO and has
been observed, among others, in cobaltites,\cite{Wu2003,Phan2004}
manganites,\cite{Salamon2001,Laiho2001} nickelates
\cite{Lander91,Freeman2006} and
cuprates.\cite{Chou95,Niedermayer98} This can be viewed as a
natural consequence of a disorder present in such complex systems
due to their chemical doping with heterovalent
cations.\cite{Dagotto2005,Hasselmann2004,Alloul2009}

In the phase diagram of cuprates, the region displaying the SG
features separates the antiferromagnetic (AF) and superconducting
(SC) regions and its characteristics are observed even inside the
SC dome.\cite{Niedermayer98,Julien2003,Mitrovic2008} Glassines of
spins, accompanied by \textit{charge} glass-like behavior, seems
to emerge with the first added
holes.\cite{Panagopoulos2005,Rajcevic2008} At low temperatures,
these holes localize, presumably at oxygen sites, and form a local
singlet on the Cu sites.\cite{Zhang88} The magnetic frustration
may be understood as a result of local ferromagnetic exchange
coupling between Cu$^{2+}$ ions in the background of
three-dimensional (3D) long-range AF order.\cite{Aharony88} The
specific role of holes, including their self-organization in the
stripes, and the details of the AF order, including the possible
existence of the spin spirals, are still under
debate.\cite{Kivelson2003,Hasselmann2004,Luscher2006} The early
observations of the spin dynamics and its evolution with doping
have indicated that the entities undergoing the SG transition are
the finite-size AF domains, cooperatively freezing in the
inhomogeneous but magnetically ordered phase.
\cite{Cho92,Niedermayer98} Thus this phase is often refereed to as
a cluster SG. While the thermodynamical SG-type behavior here is
well established, the microscopic origin is still
discussed.\cite{Chou95,Sanna2010}

The essential nature of cuprates is believed to be governed by the
physics of the Cu-O planes, which are sometimes regarded as ideal
2D Heisenberg spin systems. However, any practical realization of
such idealized layers is connected with introducing some
inevitable, \textit{internal},
disorder.\cite{Alloul2009,Coneri2010} Although the sources of such
a disorder can be cuprate-family specific (such as, for example,
the tilted CuO$_{6}$ octaedra and the stripe structure around
$x$=0.12 in La$_{2-x}$Ba$_{x}$CuO$_{4}$), the random distribution
of Coulomb impurity potentials in the layers separating the Cu-O
planes is the most common origin. Introduction of heterovalent
dopants or extra oxygen atoms is necessary to change and control
the charge-carriers concentration in cuprates but creates a
modification of the Coulomb potential that disrupts the lattice
periodicity and thus serves as an additional scattering center for
carriers in the Cu-O planes.\cite{Alloul2009}

It is natural that the above-mentioned internal Coulomb-driven
disorder is minimal in these compounds in which the carrier
reservoirs are located far from the Cu-O layers. For example, the
compounds such as HgBa$_{2}$CuO$_{4}$ and
YBa$_{2}$Cu$_{3}$O$_{6+y}$ (YBCO) are close to the cuprate clean
limit.\cite{Dagotto2005,Yang2009,Coneri2010} Doping with Ca allows
to move the YBCO apart from the clean limit in a controlled way.
The recent muon-spin-rotation ($\mu$SR) experiments on this system
give the arguments against the picture of the cluster SG phase
introduced by disorder and suggest the common ground state, named
\textit{frozen antiferromagnet}(FAF), both for low-doping regime
(i.e. in the re-entrant AF phase) and for larger dopings for which
clusters of spins coexist with percolating
superconductivity.\cite{Sanna2010,Coneri2010} The study shows that
the local-field distribution is narrow, in the strong contrast to
the canonical SG in which the distribution width is comparable to
the mean value. The disorder does not modify significantly the
hole-concentration dependence of the FAF transition temperature.
The origin of FAF remains unknown but dilution of magnetic
moments, not frustration, seems to play the crucial
role.\cite{Coneri2010}

The situation in the dirty cuprates, such as
La$_{2-x}$Sr$_{x}$CuO$_{4}$ (LSCO), is even more complicated
because of the intrinsic disorder. In LSCO, in contrast to the
clean-limit compounds, the cluster SG "order" governs solely the
physics of the system in a quite large region of the phase
diagram. The hole-doping dependence of irreversibility in
magnetization, which is found in LSCO, has been interpreted in
terms of a generic quantum glass
transition,\cite{Panagopoulos2002,Panagopoulos2004,Panagopoulos2005}
it has also been suggested to be linked to the presence of the SC
correlations above $T_{c}$,\cite{Freeman2006} and, recently, has
been re-interpreted as an argument towards existence of
spin-density-wave quantum critical point.\cite{Sachdev2010} The
NMR measurements have revealed the enhancement of spin-freezing
temperature, $T_{g}$, deeply in the SC phase of LSCO at
$x$$\simeq$0.12 and insensitiveness of this phenomenon to 1\% of
atomic disorder.\cite{Mitrovic2008} Whether the physics underlying
the cluster SG phase in the nonsuperconducting region of LSCO
phase diagram and that at $x$$\simeq$0.12 is the same, is an open
problem. The observation of \textit{charge} glass-like behavior in
lightly doped LSCO rises a question of how this dynamic charge
order evolves into SC state with increasing
$x$.\cite{Rajcevic2008} Clearly, the interplay between
superconductivity and magnetic and electronic glassiness in LSCO
is far from being well understood.

Impurities intentionally introduced into the Cu-O planes have been
widely employed to probe the properties of cuprates, with the hope
that the response of the system reveals the generic features of
the pure compound. So far, such a situation close to being perfect
has been found only in the 1D correlated systems. The magnitude of
magnetization induced in quasi-1D AF spin chain Y$_{2}$BaNiO$_{5}$
by different nonmagnetic impurities appeared to decay
exponentially, with the correlation length $\xi_{imp}$ equal to
the numerical prediction for the spin-spin $\xi$ in the pure
system.\cite{Alloul2009} In quasi-2D cuprates, the spinless
defects in the Cu-O planes induce paramagnetic moments on the
surroundings Cu ions and disorder driven by this local magnetism
extents the SG region in the phase diagram in comparison to the
pure system. As a result, the phase diagram of Zn-doped
clean-limit YBCO becomes similar to that of non-doped dirty-limit
LSCO.\cite{Mendels94} This may be viewed as an argument that
intrinsic or extrinsic - magnetically driven in this case -
disorder influence the system in the similar way.\cite{Alloul2009}

In the context of the above, Ni is an unique dopant in cuprates.
This results from its ability to built NiO$_{6}$ octaedra -
increasing the substitution of Ni ions into the Cu-sites in LSCO
leads eventually to isostructural nickelate
La$_{2-x}$Sr$_{x}$NiO$_{4}$ (LSNO). Although superconductivity is
not observed in LSNO, its phase diagram is still very reach. The
static 1D charge ordering in the form of stripes, predicted in the
context of high-T$_{C}$ superconductors (HTSC)
\cite{Zaanen89,Machida89,Schultz89} has been first observed in
LSNO\cite{Hayden92} and only later in Nd-doped
LSCO.\cite{Tranquada95} The proximity of the charge ordering
region to the SG phase suggests that they may be
related.\cite{Freeman2004,Freeman2006} Thus, a systematic
examination of the evolution of the LSCO properties with the Ni
doping may shed some light onto the problem of possible
correlation between superconductivity, stripe order and the SG
phase in HTSC.

A remarkable evolution of views on the role of Ni dopant in
cuprates has taken place, owing to accumulation of new
experimental facts.  Although nominally magnetic (3d$^{8}$, S=1),
Ni$^{2+}$ ion appears to have weaker effect on superconductivity
than the nominally non-magnetic Zn$^{2+}$ (3d$^{10}$, S=0)
ion.\cite{Pines97,Mendels99,Kofu2005} The dc susceptibility,
$\chi$, measurements have revealed very small (0.6~$\mu_{B}$)
paramagnetic moment of Ni introduced into the Cu-O
planes.\cite{Xiao90} It has been suggested that at small
concentration Ni is substituted as Ni$^{3+}$ ion.\cite{Nakano98}
The normal-state electrical transport experiments have suggested
that the quasiparticle scattering at the Ni impurity has
predominantly non-magnetic character.\cite{Hudson2001} On the
other hand, the $c$-axis optical conductivity measurements in
underdoped NdBa$_{2}$Cu$_{3}$O$_{6.8}$ show strong enhancement of
the normal-state pseudogap energy by the Ni doping while doping
with Zn is found to suppress the pseudogap. This different impact
on pseudogap has been attributed to the magnetic character of the
Ni dopant.\cite{Pimenov2005}

Just recently, it has been claimed that the Ni does not disturb
the AF spin-1/2 network in the Cu-O planes when its concentration,
$y$, is smaller than the hole concentration, $p$, in the
system.\cite{Hiraka2009} The careful measurements of the local
distortions around the Ni ions replacing Cu ions in
La$_{2-x}$Sr$_{x}$Cu$_{1-y}$Ni$_{y}$O$_{4}$ suggest that for small
concentrations, $y\leq p$ (where $p$ is equal to Sr content, $x$),
Ni serves only as a hole-absorber and creates a strongly hole-bond
state, called Zhang-Rice doublet,\cite{Zhang88} with the effective
moment S=1/2 that couples AF with S=1/2 moments of the surrounding
Cu ions. These measurements have been carried out on the single
crystals in a wide $p$-range ($p$=$x$, 0$\leq$x$\leq$0.15) but
within limited Ni concentration ($y$$\leq$0.07). Based on them, a
magnetic-impurity picture for Ni dopant in superconducting
cuprates has been completely disqualified, at least below the
optimally doped regions.\cite{Hiraka2009}

Our study shows that the actual situation is much more
complicated. We have carried out the dynamic and static magnetic
measurements of polycrystalline samples of
La$_{1.85}$Sr$_{0.15}$CuO$_{4}$ (LSCO15) doped with Ni up to large
concentration $y$=0.63, exceeding $p$ (equal to $x$=0.15 holes per
CuO$_{2}$ plane) over four times. We have found the SG behavior in
all non-superconducting samples, even in these with small $y$,
below $p$. The transition temperature $T_{g}$, when extrapolated
inside the SC region, takes finite values and approaches zero in
the $y$=0 limit. Thus, the magnetic role of Ni should not be
neglected even for $y<p$. Despite the microscopic mechanism, the
low-temperature phase of Ni-doped LSCO exhibits all features
characteristic for the SG systems: irreversibility in the
$T$-dependence of $\chi$, a peak in the zero-field cooling (ZFC)
branch of the $\chi(T)$ curve accompanied by a step in the
imaginary part of ac-susceptibility $\chi_{ac}$, the moderate
(typical for the cluster SG) frequency dependence of the peak
position in the real part of $\chi_{ac}$, and the scaling
behavior.

\section{Experimental details}
\label{sec:exp} The policrystalline samples of
La$_{1.85}$Sr$_{0.15}$Cu$_{1-y}$Ni$_{y}$O$_{4}$ (LSCNO) with
0$\leq$y$\leq$0.63 were synthesized from 4N-5N pure
La$_{2}$O$_{3}$, SrCO$_{3}$, CuO and NiO by using conventional
solid-state reaction method. The stoichiometric amounts of the
powders were carefully mixed, pressed into pellets and sintered in
a pure oxygen gas flow at 1320 K for 48 hours. After cooling down
to room temperature with the rate 2 K/min, the samples were
reground and the whole procedure was repeated two times.

The X-ray diffraction measurements were carried out at the
Bragg-Brentano diffractometer (X´Pert Pro Alpha1 MPD, Panalytical,
with a setting described in Ref.~[\onlinecite{Paszkowicz2005}]).
For determination of absolute value of unit cell parameters, NIST
SRM 676 Alumina was used as an internal reference material.
Crystallographic characterization and structure refinement was
done with help of FullProf.2k program.\cite{Rodriguez2001}

The magnetic susceptibility measurements were carried out at two
setups: the commercial SQUID magnetometer (MPMS, Quantum Design)
working in the temperature range 2 K - 400 K  and field up to 5 T
(dc and ac measurements) and the home-build setup based on
Cryogenics Consultants SQUID sensor and working in the temperature
1.6 K - 280 K and magnetic field up to 0.3 T (a part of the dc
measurements).

\section{Results}
\subsection{Crystallographic analysis}
\label{sec:crystal} 
\begin{figure}
\includegraphics[width=0.50\textwidth, trim= 5 5 5 5]
{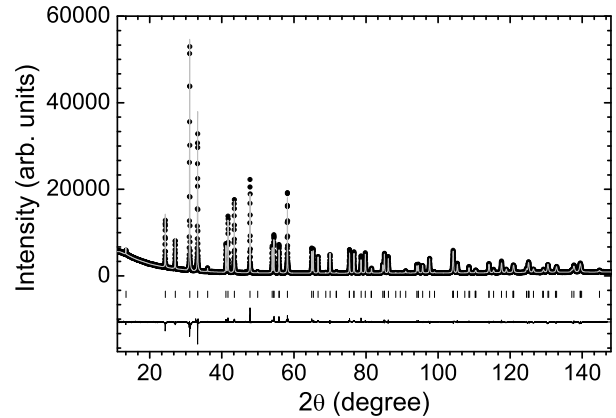}\caption{\label{Rietv} The x-ray
powder-diffraction pattern of
La$_{1.85}$Sr$_{0.15}$Cu$_{0.81}$Ni$_{0.19}$O$_{4}$. The observed
pattern is denoted by dots while the calculated one is marked as
solid line overlaying them. The bottom curve shows the difference
between them. The short vertical lines mark the positions of Bragg
reflections.}
\end{figure}

Structural and phase analysis has shown that
all samples have tetragonal K$_{2}$NiF$_{4}$-type structure (space
group I4/mmm) and include no impurity phases. An example of
Rietveld refinement plot is shown in Fig.~\ref{Rietv} for LSCNO
with the Ni content y=0.19.

\begin{figure}
\includegraphics[width=0.50\textwidth, trim= 5 5 5 5]
{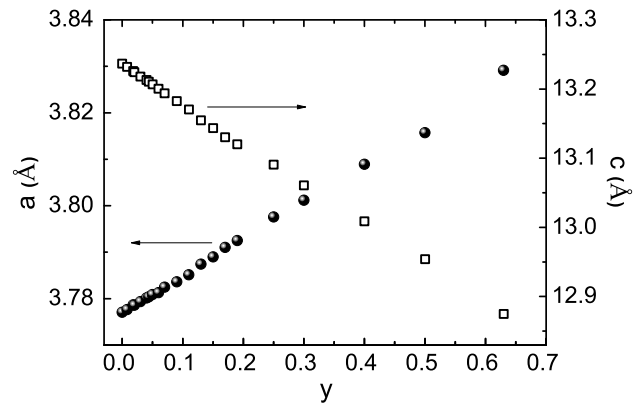}\caption{\label{lattice} Lattice
constants of LSCNO at room temperature as a function of Ni
doping.}
\end{figure}

The lattice parameters of the undoped La$_{1.85}$Sr$_{0.15}$CuO$_{4}$
sample are a=3.77709(8) \AA~and c=13.2368(4) \AA; thus, the axial
ratio c/a is equal to 3.5045(2). The powder diffraction
patterns indicate absence of any structural phase transition
in the whole investigated range of the Ni content.
The lattice constants change linearly with increasing $y$: $c$
decreases with the rate d$c/\text{d}y=-5.9(1)\times10^{-3}$
\AA~($\approx$0.02\% $c$)~per 1 at.\% of Ni while $a$ increases
with the rate d$a/\text{d}y= 8.1(1)\times10^{-4}$
\AA~($\approx$0.05\% $a$) per 1 at.\% of Ni (see
Fig.~\ref{lattice}). As a result, the $c/a$ ratio decreases with
the rate about 0.07\% per 1 at.\% of Ni. These results are
consistent with the results of the earlier studies for a large
Ni-doping range,\cite{Wu96,Zhiqiang98} but show a
considerably smaller scatter of the experimental points. The
d$c/\text{d}y$ and d$a/\text{d}y$ rates found here for $y\leq0.63$
are comparable to the rates found for $y\leq0.06$ in the
detailed X-ray-absorption fine-structure (XAFS) and X-ray powder
diffraction (XPD) measurements reported in Ref.~[\onlinecite{Haskel2001}].

Free coordinates of the atoms in the tetragonal unit cell of LSCNO
have been determined with the use of the Rietveld refinement
procedure. Position (0,0,z) of the apical oxygen atom O(2) as a
function of Ni content is discussed in Appendix \ref{sec:apical}.
Unlike the XAFS technique, XPD is not atom-specific in the sense
that only \textit{averaged} interatomic Cu/Ni-O(2) distance can be
obtained. However, the detailed analysis (see Appendix
\ref{sec:apical}) suggests that Ni-O(2) distance may change
between $y$=0.07 and $y$=1.

The bond length between the Ni atom and the in-plane oxygen atom
O(1), Ni-O(1), has been found by XAFS measurements to have two
distinct values, depending on the ratio of the hole concentration
to the Ni content.\cite{Hiraka2009} As mentioned, the XPD
technique is not capable of revealing such atom-specific details.
Since the O(1) position, ($\frac{1}{2}$,0,0), is constant in
K$_{2}$NiF$_{4}$-like structure of LSCNO, the \textit{averaged}
value of Cu/Ni-O(1) distance, resulting from the XPD analysis,
increases with $y$ in proportion to lattice parameter $a$, i.e.
with the same rate $\Delta a/(a\Delta y)$. This means contraction
of Cu(Ni)O$_{6}$ octahedra along the $c$ axis with increasing Ni
content, accompanying by simultaneous expansion in the $ab$ plane.
The resulting volume change in the whole doping range is very
small, of the order of 0.1\%. More details on analysis procedure,
free atomic positions and bond distance will be given
elsewhere.\cite{Minikayev2010}

\subsection{Normal-state dc susceptibility}
\label{dc} The measurements of $\chi(T)$ at 10 Oe reveal that the
superconductivity survives in LSCNO up to y=0.054.
\begin{figure}
\includegraphics[width=0.50\textwidth, trim= 5 5 5 5]
{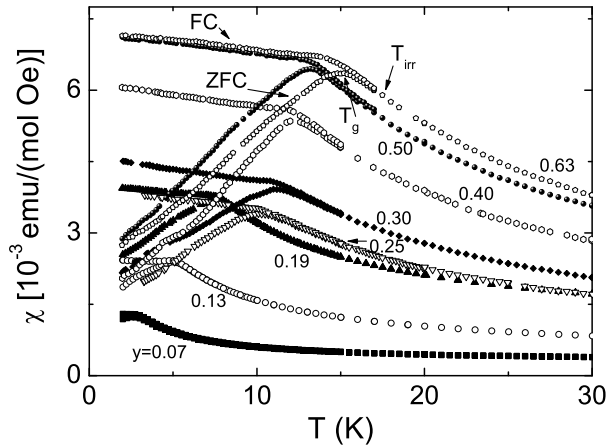}\caption{\label{ZFCandFC}The low-temperature
dc-susceptibility of LSCNO measured at 10 Oe for various Ni
content, $y$. As an example, the ZFC and FC branches of $\chi(T)$
curve and the characteristic temperatures, $T_{g}$ and $T_{irr}$,
are indicated for $y$=0.63 sample by arrows.}
\end{figure}
Above this Ni concentration, the normal-state $\chi$ vs $T$ curves
display bifurcation below a characteristic temperature, $T_{irr}$,
depending on the thermal-magnetic history of the sample, i.e.
whether it was cooled down in the zero magnetic field (ZFC mode)
or in the non-zero field (FC mode). The representative curves are
shown in Fig.~\ref{ZFCandFC}. The ZFC branch has a rounded but
very well defined peak, position of which, $T_{g}$, is slightly
below $T_{irr}$. Let us note that in the canonical SG usually an
opposite situation is observed, i.e. $T_{irr}\lesssim T_{g}$. The
behavior observed in LSCNO can be a manifestation of the existence of
magnetic clusters in the system.\cite{Marcano2007}

\begin{figure}
\includegraphics[width=0.50\textwidth, trim= 5 5 5 5]
{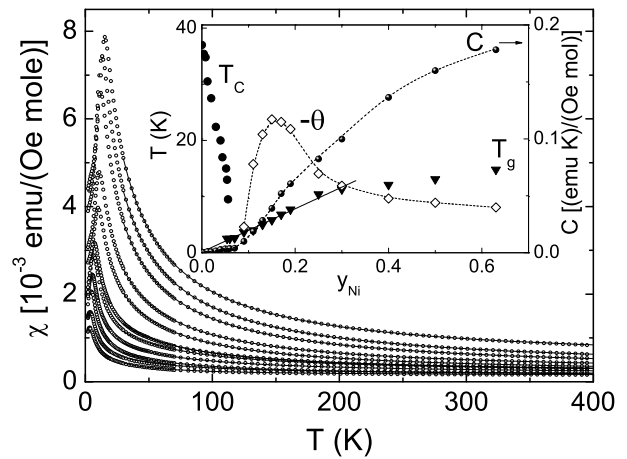}\caption{\label{CW} The dc-susceptibility
of LSCNO, measured at 1 kOe field, for Ni content, $y$, equal to
(from bottom up) 0.09, 0.11, 0.13, 0.15, 0.17, 0.19, 0.25, 0.30,
0.50 and 0.63. The solid lines are the best fits to the
Curie-Weiss law in the 30 K - 400 K range. Inset shows the
characteristic temperatures $T_{C}$ and $T_{g}$ and effective
fitting parameter $\theta$ (left scale) as a function of $y$. The
solid line is the linear fit to $T_{g}$ vs $y$ dependence for
$y\leq0.30$. In addition, the fitting parameter C is shown as a
function of $y$ in the whole doping range (right scale). The
dashed lines are guides to the eye.}
\end{figure}

The temperature of the peak in the ZFC $\chi(T)$ curve, $T_{g}$,
which we take as the temperature of a transition to the SG phase,
increases linearly with $y$ up to $y$=0.30 (see the inset to
Fig.~\ref{CW}). The smallest concentration $y$, for which we
observe the bifurcation of $\chi(T)$ curve, is equal to 0.056. The
temperature $T_{g}$, when extrapolated linearly outside our
temperature measurement window and inside the SC phase, has finite
values and decreases to zero at $y$=0 (see the solid line in
the inset to Fig.~\ref{CW}). This strongly suggests that Ni dopant
in the sublattice of surrounding Cu spins displays a magnetic
nature starting from the smallest concentration $y$.

It should be emphasized that the deviation of $\chi(T)$ measured
in the FC mode from that measured in the ZFC mode cannot alone
prove that a SG state develops below the bifurcation temperature
of $\chi(T)$ curve. The superparamagnetic systems and conventional
ferromagnets with a broad distribution of potential barriers
exhibit such a feature as well. In addition, the ZFC-FC splitting
of $\chi(T)$ is observed in disordered AF compounds that may be
more relevant to the LSCNO system.\cite{Korner2000} Thus, in the
next sections, we will present the additional characteristics
confirming the transition to the SG phase: a logarithmic frequency
dependence of the real part of ac susceptibility, $\chi'$, below
the transition, a step in the imaginary part, $\chi''$, a small
frequency dependence of the peak position in $\chi'(T)$ curve,
described by the standard critical slowing-down formula, time
decay of the thermoremnant magnetization and, finally and the most
conclusive, the scaling behavior of the nonlinear
dc-susceptibility.

Before that, however, we will shortly describe the $\chi$ behavior
at temperatures well above $T_{g}$. The normal-state $\chi(T)$ at
larger temperatures for Ni concentration below $y$=0.09 can be
analyzed in terms of the universal empirical curve $F$, proposed
for LSCO without parametrization independently by
Johnston\cite{Johnston89}and Nakano.\cite{Nakano98} The universal
function $F$ was found to describe the effective susceptibility of
the Cu$^{2+}$ spin sublattice in LSCO with various Sr content when
the $\chi(T)$ data are represented in the reduced parameters
$(\chi-\chi_{0})/(\chi-\chi_{max})$ and $T/T_{max}$, where
$\chi_{0}$ is the $T$-independent  sum of three components: (1) -
the isotropic contribution from the closed Cu shells, (2) - the
Van Vleck contribution  and (3) - the contribution from the charge
carriers, and $\chi_{max}$ is the maximal value that $\chi(T)$
reaches at $T_{max}$. Using the universal function from
Ref.~[\onlinecite{Nakano98}], we have fitted $\chi(T)$ to the
formula $\chi(T)=\chi_{0}+A\cdot F(T/T_{max})+C/T$ and have
calculated the effective magnetic moment introduced by Ni ion,
$\mu_{eff}$, from the Curie term, $C/T=N\mu^{2}_{eff}/3k_{B}T$,
with the rest of parameters having their usual meaning (see Ref.
[\onlinecite{MalinowskiAndBezussy2010}]). The moment $\mu_{eff}$
is constant up to $y$=0.07 and equal to $(0.7\pm0.05) \mu_{B}$ per
Ni ion, what is very close to the value found previously when a
linear function instead of $F$ has been used in the
analysis.\cite{Xiao90} Note that there is no need to add any
finite paramagnetic Curie temperature, $\theta$, in the Curie term
to reproduce the experimental data well, as it was necessary for
Ni-doped LSCO with x=0.18 Sr content.\cite{Ishikawa92}

The $\chi(T)$ term described by the $F$ function is not detectable
when Ni concentration exceeds $y$=0.09. For these large $y$, the
$\chi(T)$ data above $\sim$30 K are well reproduced by the
Curie-Weiss (CW) law, with the finite \textit{negative} $\theta$,
$\chi=\chi_{0}+C/(T-\theta)$. The best fits are shown by the solid
lines in Fig.~\ref{CW}. The calculated magnetic moment $\mu_{eff}$
does not remain constant, as it does for smaller $y$, and
increases with increasing $y$ to reach about 1.6 $\mu_{B}$ per the
Ni ion for $y$=0.5. Assuming temporary that all Ni ions are in the
same magnetic state this would mean that the spin of Ni
ion is much closer to $S$=1/2 (for which $\mu_{eff}$=1.73
$\mu_{B}$) than to $S$=1 ($\mu_{eff}$=2.82$\mu_{B}$). Note that
the identical value (1.6$\pm$0.1$)\mu_{B}$ was obtained for
Ni-doped YBa$_{2}$Cu$_{3}$O$_{6+x}$, where Ni moment remains
quasi-constant with hole doping in the investigated range up to
$y$=0.04.\cite{Mendels99}

The CW term in the temperature dependence of susceptibility
indicates the presence of the localized magnetic moments in the
system. For small $y$$<$$0.09$, when the magnetism of the Cu-ions
sublattice is clearly described by the function $F$ and it is not
much altered by the Ni ions introduced in small quantities, the
Curie term in $\chi(T)$ can be fully attributed to Ni ions, all
being in the same magnetic state.\cite{Xiao90} The resulting
$\mu_{eff}=g\sqrt{S(S+1)}=0.7\mu_{B}$ means that the spin
$S$ of Ni ion, immersed in the surrounding Cu ions sublattice,
is equal to 0.11. Such a small value can be
explained by the formation of the strongly hole-bound state of Ni
ion, represented as Ni$^{2+}$\textit\underline{L}, as proposed in
Ref.~[\onlinecite{Hiraka2009}]. However, we observe the increase
of the $\mu_{eff}$ starting at $y$$\approx$0.09, not at $y$=0.15
(being equal to the hole concentration in the system),
as it is expected
from the picture suggested in Ref.~[\onlinecite{Hiraka2009}].
Abrupt increase of the $C$ in the CW law for $y$$>$0.07 suggests
that at least part of the Ni ions is in the magnetic state different
from Ni$^{2+}$\textit\underline{L}, even at concentration below
$y$=0.15. However, the macroscopic susceptibility measurements
alone do not allow to separate out the observed effective moment
into two components coming from the Ni ions in two different
magnetic states.

The values of $\theta$ for a given $y$ display some dependence on
the $T$-range used in fitting $\chi(T)$. This makes to treat
$\theta$ as the effective parameter rather than the true constant.
In addition, there might be some crystal electric field effects,
averaged in the policrystalline samples. Nevertheless, the
negative sign of the effective paramagnetic CW temperature
$\theta$  indicates that the dominant exchange interactions in the
system are antiferromagnetic. As it is shown in the inset to
Fig.~\ref{CW}, the $\theta$ exhibits strong dependence on Ni
content. Its absolute value, $|\theta|$, increases abruptly when
$y$ increases from $y$$\approx$0.09 up to $y$=0.15 This is
followed by a rapid decrease, and a regime of saturation for
$y$$>$0.30 (with a weak tendency to decrease with increasing $y$).
This behavior can be understood as a result of trapping the mobile
holes by the Ni ions and will be discussed in the
Sec.~\ref{sec:Ni}.

\subsection{Ac susceptibility}
\label{ac}
\begin{figure}
\includegraphics[width=0.50\textwidth, trim= 5 5 5 5]
{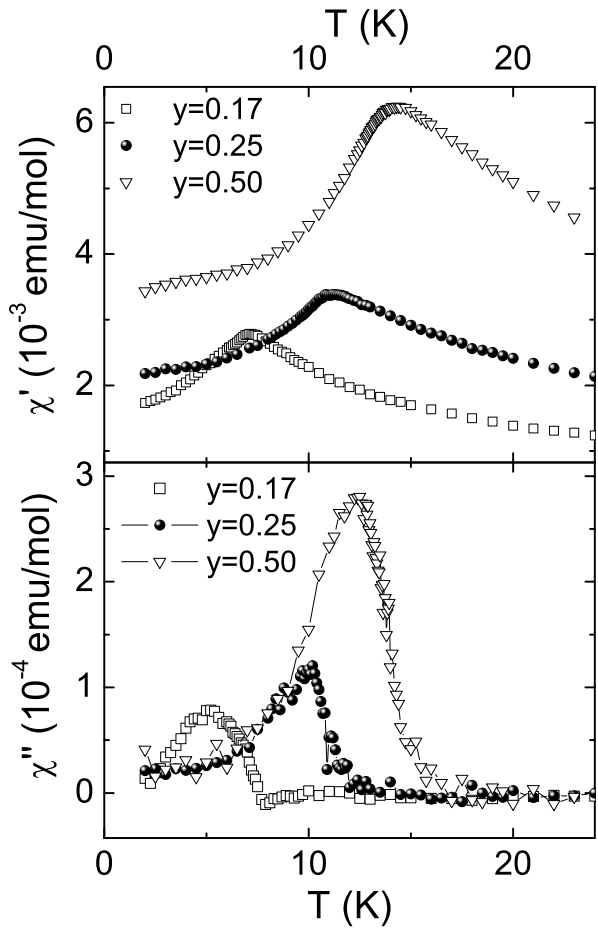}\caption{\label{3samples} The real
$\chi'$ and imaginary $\chi''$ components of the ac susceptibility
of LSCNO with y=0.17, 0.25 and 0.50 Ni content as a function of
temperature, measured at f=1 Hz and with 1 Oe amplitude of the
applied ac field and without any dc-bias.}
\end{figure}
The metastable SG state is usually characterized by ac
susceptibility, $\chi_{ac}$. In Fig.~\ref{3samples} we show
$\chi_{ac}(T)$ curves for LSCNO for several Ni contents, $y$=0.17
(LSCNO17), $y$=0.25 (LSCNO25) and $y$=0.50 (LSCNO50). A quite
sharp cusp is visible in the in-phase component, $\chi'(T)$, for
all samples. Its temperature, $T_{f}$, is always a bit larger than
the temperature of the maximum in dc-$\chi(T)$, $T_{g}$, and roughly
coincidences with the temperature of the inflection point in the step
of the out-of-phase component, $\chi''(T)$. At larger
temperatures, above $T_{f}$, $\chi''(T)$ is equal to zero, while
below $T_{f}$ has a finite value. Such a behavior is
characteristic for SG transition and allows to distinguish the SG
compounds from the disordered AF systems, in which $\chi''(T)$ is
constant and remains equal to zero even below the temperature of
the transition.\cite{Korner2000,Sullow97,Mulder82}
\begin{figure}
\includegraphics[width=0.50\textwidth, trim= 5 5 5 5]
{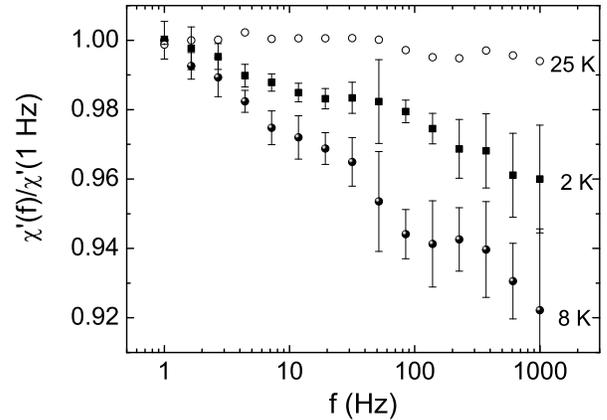}\caption{\label{flog} Normalized real part of
$\chi_{ac}$ for LSCNO25 as a function of frequency for
different temperatures - above and below $T_{g}$. The amplitude of
ac field was 1 Oe and the measurements were carried out in 10 Oe
dc field. The error bars for the frequency scan at $T$=25 K are
similar to these for the scans at 2 K and 8 K.}
\end{figure}

In Fig.\ref{flog} we depict the real part of $\chi_{ac}$,
normalized to the value at 1 Hz, as a function of frequency for
LSCNO25, as an example. At 25 K, i.e. above $T_{g}$, in the
paramagnetic state, the variation of $f$ over three decades does not
influence $\chi'$ in a noticeable way. Below $T_{g}$, in the SG
state, the $\chi'$ exhibits a logarithmic frequency dependence.
Such a frequency dependence has been predicted theoretically for
a short-range Ising SG\cite{Fisher86} and it has been observed
for many SG systems.\cite{Giot2008} However, this logarithmic
relationship is not unique for the SG, because it is
exhibited by any two-level disordered system,
dynamics of which is governed by activated processes with
a broad distribution of the activation barriers
heights.\cite{Pytte87} Note that at 8 K, closer to $T_{g}$, the
system is more sensitive to variation of frequency then at 2 K,
"deeper" in the frozen state.

The ac measurements at various frequencies reveal that the
position of the maximum in $\chi'(T)$ curve, $T_{f}$, moves toward
higher temperature and the magnitude of
$\chi'$ decreases with increasing frequency.
Such a behavior is expected for a SG system.
As a raw measure of this frequency dependence, the parameter
$\delta=d\text{ln} T_{f}(\omega)/d \omega$ is used.\cite{Binder86}
In experimental practice, a shift between two outermost frequencies
accessible in experiment is usually employed and $\delta$ is
calculated as $\delta=\Delta T_{f}/(T_{f}\Delta
\text{ln}\omega)$.\cite{Mulder82,Marcano2007} The values of
$\delta$ obtained in this way for LSCNO system do not show any
obvious correlation with $y$ and are equal to 0.012, 0.012 and
0.014 for $y$=0.17, 0.25 and 0.50, respectively.

For systems with noninteracting entities (which can be either
particles or magnetic clusters), the values of $\delta$ of the
order of 0.1 have been reported\cite{Dormann88} and such a
relatively high sensitivity to frequency is predicted in the
classical model of superparamagnetism.\cite{Neel49,Brown63} Any
interactions between particles weakens this high sensitivity and
this is why almost two orders of magnitude smaller values of
$\delta$ are measured in the canonical spin glasses (e.g.
$\delta$$\approx$0.002 for CuMn, Ref.[\onlinecite{Mulder81}])
or in the SG phase of other systems, such as manganites, where
$\delta$$\approx$0.003 is reported.\cite{Giot2008} Finally, in
the case of well-ordered ferromagnetic or AF systems with even
stronger interactions only MHz and GHz frequencies are
sufficiently large to cause any observable shift in the $\chi'(T)$
peak position.\cite{Mydosh93}

The values of $\delta\sim0.01$ obtained for LSCNO lay between
these extremes and are typical for cluster glasses, i.e. systems
with randomly distributed interacting magnetic
clusters.\cite{Li2005} Recently, $\delta=0.016$ was observed in
Ce$_{2}$CuGe$_{3}$, which is an example of so called nonmagnetic
atom-disorder SG system with possible (ferromagnetic)
clusters.\cite{Tien2000} To recapitulate, the use of the simple but
model-independent criterion based on the parameter $\delta$
strongly suggests the possibility of the existence of spin
clusters in LSCNO. The frequency dependence of $T_{f}$ will
be discussed in more detail in Sec.~\ref{sec:dynamical}

\subsection{Magnetic hysteresis}
\label{sec:hyst} 
\begin{figure}
\includegraphics[width=0.50\textwidth, trim= 5 5 5 5]
{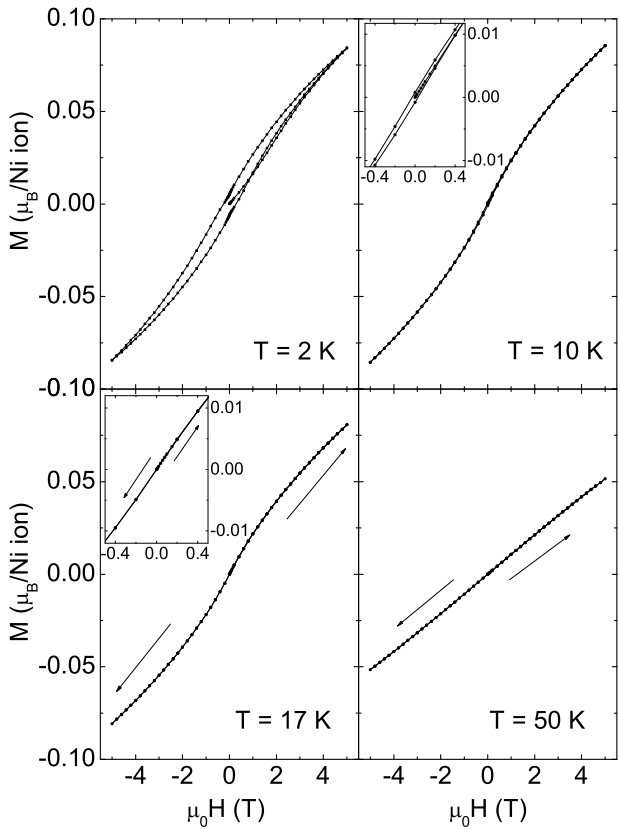}\caption{\label{hysteresis} Evolution of
the $M$ vs $H$ behavior with temperature for y=0.50 sample. The
$M-H$ isothermal curves are shown for 2 K (well below
$T_{g}$=13.1K), 10 K (still below $T_{g}$), 17 K (between $T_{g}$
and $T_{irr}$) and 50 K (well above $T_{f}$). Insets show details
at low fields at T=10 K and at T=17 K.}
\end{figure}

The presence of spin clusters should be reflected
in the magnetic hysteresis. Evolution of the magnetization loops
in LSCNO50 is illustrated in Fig.\ref{hysteresis}, where the
isothermal $M(H)$ curves are depicted at several temperatures.
The curves for LSCNO17 and LSCNO25 have qualitatively similar
features. No trace of saturation is visible even at the lowest
$T$. At 2 K, the individual branches of the $M(H)$ curve obtained
after ZFC have a characteristic "S" shape - the initial slope of
the curve is smaller than the slope at inflection point at nonzero
field. This is particularly clearly visible in the virgin curve.
Such "S" shape of $M$ vs $H$ curve is typical for SG systems in a
frozen state.\cite{Binder86} With increasing temperature the
S-shaped curve smoothly evolves into a straight line, indicating
the paramagnetic behavior at high $T$. Hysteresis is observed
only below $T_{g}$. The coercive field, taken from the hysteresis
loops measured at 2 K, increases linearly with $y$ and is equal to
1.0, 1.4 and 2.6 kOe for y=0.17, 0.25 and 0.50, respectively. A
high-field part of the virgin curve (above 9 kOe for y=0.50 at 2
K) lies outside the hysteresis loop. This suggests the presence of
metastable states in the system.

The existence of a hysteresis loop clearly excludes
superparamagnetism as a candidate for the ground magnetic state of
LSCNO, since superparamagnetism is a thermal equilibrium
behavior.\cite{Bean59} In the case of ferromagnetic or canted AF
clusters existing in the system, a hysteresis is expected but with
large initial susceptibility because clusters are at first
saturated along their local easy axis and only after that various
clusters became fully aligned along the applied
field.\cite{Marcano2007} If clusters reverse their magnetization
coherently then hysteresis loops have almost rectangular
shapes.\cite{Westerholt86} In LSCNO just the opposite behavior is
observed - the initial $\chi$ is smaller then this at inflection
point in $M(H)$ curve. Thus, it is reasonable to conclude that
$M(H)$ curves do not show any feature indicative of (large)
clusters in LSCNO. The $M(H)$ curves remain non-linear for
$T_{g}<T<T_{irr}$ indicating that SG phase starts to built over a
broad T-range. For LSCNO25, nonlinearity in the $M$ vs $H$
dependence is observed even at 25 K, i.e. at $T\simeq2.5T_{g}$,
well above $T_{irr}$. The deviations of the $M(T)$ curve
measured at constant field from the CW dependence fitted
in a wide $T$-range (up to 400 K), appear at around 30 K.
These features indicate that short-range AF correlations
and fluctuations exist well above $T_{g}$.
Such precursors of cooperative freezing at $T_{g}$ have been
observed in metallic SG with $\theta$$>$0 at temperatures higher
than 5$T_{g}$\cite{Morgownik81} and in the amorphous SG with both
signs of $\theta$ at even higher temperatures
$T$$\approx$$12-20T_{g}$.\cite{Morgownik82,Morgownik84,Rao83}

\subsection{Decay of thermoremnant magnetization}
\label{remnant} Since the SG system in a frozen state
can react to the applied field slowly, the magnetization
curves obtained after ZFC procedure do not give information about
the thermal equilibrium of SG but rather reflect a slow increase
of magnetization (and thus susceptibility) with
time.\cite{Binder86} The underlying physics is similar to the
one that governs the decay of the remanent magnetization with time.

To investigate the mechanism by which the system decays back to
equilibrium we have applied a following procedure. First,
the magnetic field of 1000 Oe has been turned on at 200 K.
Subsequently, the sample has been cooled down to 200 K
during 80 min, then the temperature has been kept constant for
10 min, and, finally, the magnetic field has been switched off.
Next, the remnant magnetization has been measured vs time
at $T=2$ K, starting immediately after the field became zero.
\begin{figure}
\includegraphics[width=0.45\textwidth, trim= 5 5 5 5]
{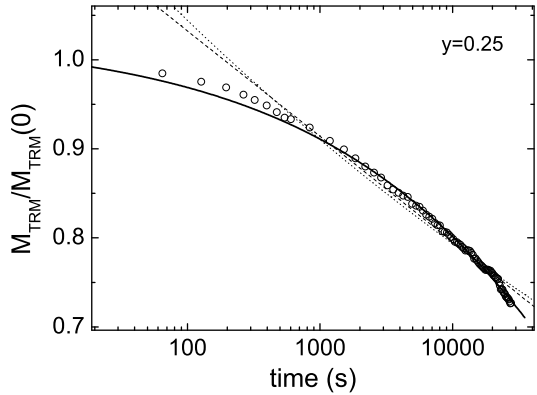}\caption{\label{decay} Time dependence of the
thermoremanent magnetization $M_{TRM}$ for the LSCNO sample with
y=0.25. $M_{TRM}$ is normalized to the value measured immediately
after the field is set to zero, $M_{TRM}(0)$. The thick solid line
is the best fits to a stretched exponential dependence, the dotted
thin line - to Eq.~(\ref{PwLaw}) and the dashed one - to
Eq.~(\ref{log}).}
\end{figure}
As it can be seen in Fig.\ref{decay}, the time decay of this FC
thermoremanent magnetization, $M_{\text{TRM}}$, is well described
by a stretched exponential formula,
\begin{equation}
M_{TRM}(t)=M_{0}\text{exp}[-(t/\tau)^{1-n}].\label{stretched}
\end{equation}

This form is commonly used to describe different relaxation
phenomena, including magnetic, optical and mechanical ones, in
different complex random systems with a distribution of relaxation
times.\cite{Degiorgio90,StretchedExponential} However, no
generally accepted microscopic explanation of this behavior exists
so far. Percolation model puts the restrictions on the possible
values of stretching exponents, $1/3\leq1-n\leq1$.\cite{Lois2009}
The values of $1-n$ below the mean-field value, 1/3, were
reported, albeit they are believed to be
non-intrinsic\cite{Phillips96} or related to fragility of some
glass formers.\cite{Bohmer93} The theory predicts that presence
of the long-range forces (in addition to the obvious short-range
forces) modifies the glassy relaxation dynamics, reducing
$1-n$ (within the limits given above).\cite{Phillips2006,Jund2001}
The relationship (\ref{stretched}) was theoretically predicted also
for the SG systems.\cite{Continentino86,Ogielski85,Dominicis85}
The numerous experimental data on the canonical SG, such as,
for example, Ag:Mn,\cite{Hoogerbeets85} confirm validity of
this formula for the description of the relaxation of magnetization
in these systems.\cite{Binder86}
The best fit to Eq.~(\ref{stretched}) for
LSCNO25 is visible in Fig.~\ref{decay} as the solid line.

We have also tried to fit the data to other formulas used to
describe decay of magnetization, a power law and a logarithmical
dependence,
\begin{equation}
M_{\text{TRM}}(t)=M_{0}t^{-\gamma},\label{PwLaw}
\end{equation}
\begin{equation}
M_{\text{TRM}}(t)=M_{0}-S\text{ln}(t),\label{log}
\end{equation}
respectively.\cite{Sinha96} Eq.~(\ref{PwLaw}) has been used to
describe relaxation of magnetization in the systems with both
long-range-AF and long-range-ferromagnetic
order.\cite{Sinha96,Patel2002} Numeric simulations predict that
this equation should be also applicable to the SG
systems.\cite{Kinzel79} Eq.~(\ref{log}) has been found to be valid
for the systems where the energy barriers over which magnetic
relaxation takes place are uniformly distributed from zero to a
certain maximal energy and the behavior consistent with this
equation has been observed in several different SG systems.
\cite{Guy77,Guy78,Dhar2003} It has been noted that when the decay
parameter $\gamma$ is small, the experimental data in the limited
time interval can be fitted both by Eq.~(\ref{PwLaw}) and
Eq.~(\ref{log}) equally well.\cite{Kinzel79,Smith94} In all these
equations (\ref{stretched}-\ref{log}) $M_{0}$ is one of the
fitting parameters but, as pointed out in
Ref.[\onlinecite{Tiwari2005}], in Eq.~(\ref{PwLaw}) and
Eq.~(\ref{log}) it depends on the time unit used (since this is a
fitted value of $M_{TMR}$ at $t$=1). The dotted line presented in
Fig.\ref{decay} is the best fit to Eq.~(\ref{PwLaw}), while the
dashed one is the best fit to Eq.~(\ref{log}). The magnetization
is normalized by $M_{TRM}(0)$, the \textit{measured} value of
$M_{TRM}$ immediately after the field is set to zero. The
measurement was sufficiently long ($\sim 8$ hours) to distinguish
between the different possible functional forms, given by
Eqs.~(\ref{stretched})-(\ref{log}). The quality of the best fits
to the functions (\ref{PwLaw})-(\ref{log}) is not satisfactory, as
it can be easily seen in Fig.~\ref{decay}.

The calculated values of $\chi^{2}$ and correlation coefficient
$R^{2}$ in the standard analysis unambiguously show that the
stretched exponential form (\ref{stretched}) is the best
description of magnetization decay in LSCNO.  The best fit to
Eq.~(\ref{stretched}) gives $1-n=0.32\pm0.01$. The
parameter $n$ is in perfect agreement with theoretical predictions
for the SG systems,\cite{Campbell86,Campbell88} with the simulations
on a 3D Ising SG\cite{Ogielski85} and with the experimental
results on the canonical SG, which also
give $1-n\approx1/3$.\cite{Chamberlin84}

\section{Discussion}

\subsection{Static scaling}
\label{sec:static}The definitive feature that corroborates the
presence of the SG phase in LSCNO is the scaling behavior.
Introducing the SG susceptibility, $\chi_{SG}$, proportional to
averaged (thermally and spatially)\cite{Fischer75,Sherrington75a}
square of the spin correlation function,
$\chi_{SG}\propto\displaystyle\sum_{i,j}[\langle S_{i}\cdot
S_{j}\rangle^{2}_{T}]_{av}$, allows to analyze the SG transition
within the framework of the second-order phase transition theory,
with diverging correlations at $T_{g}$ and onset of the "order"
below $T_{g}$, and next to implement the static scaling
hypothesis\cite{Widom65, Kadanoff66} to description of the
SG.\cite{Fischer91}

Experimentally, the $\chi_{SG}$ is measurable through the
dimensionless nonlinear susceptibility, defined as
$\chi_{nl}(T,H)\equiv1-M(T,H)/\chi_{l}H$.\cite{Chalupa77,
Suzuki77,Omari83} The linear susceptibility, $\chi_{l}$,
comes from the measurements at low field. This definition of $\chi_{nl}$
means that the measured temperature-dependent susceptibility,
$\chi\equiv M/H$, can be written as
$\chi$=$\chi_{l}(1-\chi_{nl})$. Note that sometimes the term
"nonlinear susceptibility" refers in the literature to the
coefficient at the third power of field in the expansion of
magnetization in odd powers of field, $\partial^{3}M/\partial
H^{3}|_{H=0}$. In the following, by $\chi_{nl}$ we mean the
"whole" nonlinear part of $\chi$.\cite{Dekker88} To use
$\chi_{nl}$ in the scaling analysis, one needs to stand up to the
problem with estimation of the critical region where the scaling
should hold and to the fact that the state below the SG transition
temperature is not a state of thermodynamic
equilibrium.\cite{Fischer91_ch8} Despite these difficulties,
$\chi_{nl}$ has been found to be a good tool to investigate a
possible thermodynamic phase transition in the various 2D and 3D
SG systems.\cite{Beauvillain84}

\begin{figure}
\includegraphics[width=0.5\textwidth, trim= 5 5 5 5]
{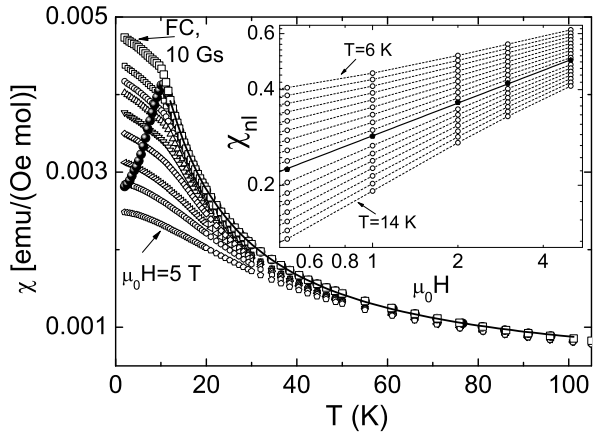}\caption{\label{Fields} Magnetic
susceptibility of LSCNO25 as a function of temperature at various
fields. The large symbols denote ZFC (black spheres) and FC (open
squares) data measured at $\mu_{0}H$=10 Gs. The solid line is the
best fit of CW function (plus constant) to the FC data above 12 K.
The small open symbols represent data measured at 0.02, 0.1, 0.2,
0.5, 1, 2, 3 and 5 T (from top to bottom) in the FC mode. In the
inset, the open circles denote the nonlinear susceptibility
$\chi_{nl}$ as a function of field at various temperatures at 0.5
K intervals, from 6 K to 10 K and from 11 K to 14 K (from top to
bottom). The solid line is the best fit of the dependence
(\ref{delta}) (see text) to the data at $T_{g}$=10.4 K (denoted as
solid circles). The dashed lines are the guides to the eye.}
\end{figure}

In Fig.\ref{Fields} we depicted the susceptibility of LSCNO25
measured at various fields from $\mu_{0}H$=10 Gs to 5 T.
The temperature-dependent part of the susceptibility at low field
and at temperatures above $\approx$1.2 $T_{g}$ is well described
by the CW dependence, $\chi(T)-\chi_{0}=C/(T-\theta)$. The best
fit to the FC $\chi(T)$ data in the range 12 K - 100 K, showed in
Fig.\ref{Fields} as the solid line, gives negative effective
$\theta$=$-5.3\pm0.1$ K, indicating the presence of the AF
correlations. We will take only the $T$-dependent part of the
measured $\chi$ into account in the following scaling analysis. We
assume that $\chi(T)$ measured at 10 Oe is a good approximation of
$\chi_{l}$ and thus $\chi_{l}=C/(T-\theta)$. At low temperatures,
the magnetization increases nonlinearly with the field and the
$\chi(T)$ deviates from the CW function towards the smaller values
at larger fields, as it is clearly visible in Fig.\ref{Fields}.

The scaling theory predicts for $\chi_{nl}$ the relationship
\begin{equation}
\frac{\chi_{nl}}{\mid
t\mid^{\beta}}=f_{\mp}\left(\frac{H^{2}}{|t|^{\beta+\gamma}}\right)
\label{scal1},
\end{equation}
where $t$ is the reduced temperature, $t$=$(T-T_{g})/T_{g}$,
$\beta$ and $\gamma$ are the critical exponents, and $f_{-}$
($f_{+}$) is the scaling function for $t<0$
($t>0$).\cite{Suzuki77,Barbara81,Malozemoff83} (Another approach
to the static scaling, sometimes used in the literature but
clearly giving unsatisfactory results in the case of LSCNO, is
discussed in Appendix \ref{AnotherScaling}.) The scaling functions
$f_{\mp}(x)$ behave as const$\cdot x^{2/\delta}$ in the large-$x$
limit.\cite{Chalupa77,Barbara81} This means that right at $T_{g}$
we have
\begin{equation}
\chi_{nl}\propto (H^{2})^{1/\delta},\label{delta}
\end{equation}
where $\delta$ is another critical exponent, related to $\beta$ and
$\gamma$ by the scaling law
\begin{equation}
(\delta-1)\beta=\gamma\label{ScalLaw}.
\end{equation}

In the inset to Fig.\ref{Fields} we present the $\chi_{nl}$ vs $H$
curves for various temperatures. They are results of the
isothermal cross-cuts of the $\chi_{nl}(T)$ curves calculated from
the experimental data for various fields. The position of the
maximum in the ZFC $\chi(T)$ curve at 10 Oe,
equal to 10.4$\pm$0.1 K, is taken as $T_{g}$.
The $\chi_{nl}(H)$ curves on log-log scale change their curvature
sign around $T_{g}$. Right at $T_{g}$, the data are described by a
linear dependence on the log-log scale and the best fit to
Eq.~(\ref{delta}), shown in the inset as the solid line, yields
$\delta$=5.8$\pm$0.1.

With this value of $\delta$ we have adjusted $\beta$ and have
calculated $\gamma$ from Eq.~(\ref{ScalLaw}) to obtain the optimum
coincidence of the data on two universal curves, one for $t<0$ and
second for $t>0$, in the $\chi_{nl}/|t|^{\beta}$ vs
$H^{2}/|t|^{\beta+\gamma}$ plot. The best qualitative collapsing
of the data to these two separate curves has been found for
$\beta$=0.75, what implies $\gamma$=3.6 (see Fig.\ref{scal1Fig} in
Appendix \ref{AnotherScaling}). We have estimated the uncertainty
of the adjusted parameter to be $\sim$0.05, i.e.
$\beta$=0.75$\pm$0.05 and thus $\gamma$=3.6$\pm$0.3.

\begin{figure}
\includegraphics[width=0.5\textwidth, trim= 5 5 5 5]
{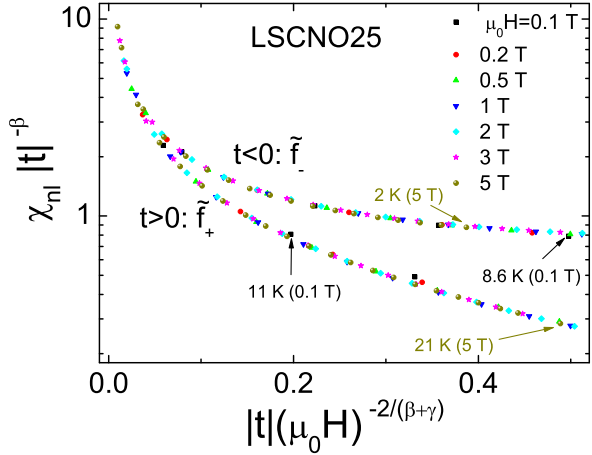}\caption{\label{scal1aFig} (Color online)
Scaling plot for LSCNO25 according to Eq.~(\ref{scal1a}) with
$\beta$=0.75 and $\gamma$=3.6 for the nonlinear susceptibility
$\chi_{nl}$ at various magnetic fields from 0.1 T to 5 T at
temperatures below ($t$$<$0) and above ($t$$>$0) the SG-transition
temperature $T_{g}$=10.2 K. The error bars are of the symbol
size.}
\end{figure}

To estimate and to visualize the critical temperature region where
the scaling is valid, it is better to use the argument of the
scaling function that is linear in $t$. Such improved form of the
scaling has been proposed by Geschwind et al. for the equation
$\chi_{nl}=H^{2\beta/(\beta+\gamma)}G(x)\equiv
H^{2\beta/(\beta+\gamma)}G(H^{2}/|t|^{\beta+\gamma})$.
\cite{Geschwind90} Raising the argument $x$ to the power of
$-1/(\beta+\gamma)$ makes it linear in $t$. In the same way,
Eq.~(\ref{scal1}) may be reformulated as
\begin{equation}
\frac{\chi_{nl}}{\mid
t\mid^{\beta}}=\tilde{f}_{\mp}\left(\frac{|t|}{H^{\frac{2}{\beta+\gamma}}}\right)
\label{scal1a}.
\end{equation}
The scaling plot with the use of this equation for $\beta$=0.75
and $\gamma$=3.6 is presented in Fig.~\ref{scal1aFig}.
In the logarithmic scale of $y$-axis  all values of
$\chi_{nl}|t|^{-\beta}$, varying over two decades, are given equal
weight and this allows to notice and compare any potential
deviations from the universal curves at different ordinates.  The
scaling validity region at lower fields is noticeably smaller than
at larger fields, i.e. the scaling does not work for the same
argument $|t|/H^{2/(\beta+\gamma)}$  of function $\tilde{f}_{\mp}$
for which the scaling at larger fields is still valid. In
terms of temperature critical region, this means that at the field
0.2 T scaling is valid in the interval $\sim0.8T_{g}-1.1T_{g}$.
The deviations of the data outside this $T$-region from the
universal curves are larger than the measurements uncertainties
(which are of the symbol size in Fig.~\ref{scal1aFig}). The
scaling region expands with increasing field and the quality of
scaling at 5 T is still excellent even at temperatures as far from
$T_{g}$ as $0.2T_{g}$ (i.e. the lower limit of our measurement
window) and $2T_{g}$. Such a large scaling region has been found
experimentally to be typical for the canonical SG
systems,\cite{Barbara81} in agreement with the numerical
simulations for the 3D Ising SG, by which the critical region has
been estimated to extend up to $|t|\approx0.55$.\cite{Ogielski85}

\begin{figure}
\includegraphics[width=0.5\textwidth, trim= 5 5 5 5]
{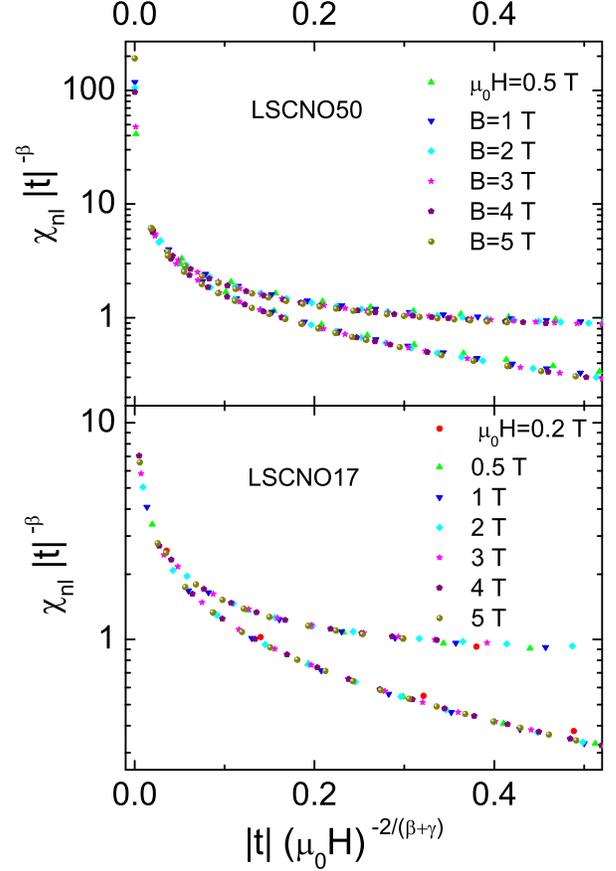}\caption{\label{Scal17and50} (Color
online) Scaling plot for the nonlinear susceptibility $\chi_{nl}$
of LSCNO17 (LSCNO50) according to Eq.~(\ref{scal1a}) with
$\beta$=0.55 ($\beta$=0.75) and $\gamma$=3.2 ($\gamma$=3.8).}
\end{figure}
Results of the identical scaling procedure for LSCNO17 and LSCNO50
are similar to those for LSCNO25. In particular, the scaling
according to Eq.~(\ref{scal1}) and its improved form,
Eq.~(\ref{scal1a}), gives clearly the better results than the
scaling according to \ Eq.~(\ref{scal2}). This remains true even
for LSCNO17 where the strongest AF correlations among the
investigated samples are expected and thus one might expect that
the scaling approach described in Appendix \ref{AnotherScaling}
would work better than that described by Eq.~(\ref{scal1a}). The
best collapsing of the data at different magnetic fields onto two
branches of the universal curve - below and above $T_{g}$ - have
been found for LSCNO17 for $\beta$=0.55, i.e. the value closer to
$\beta$=0.5, predicted by the numerical simulations for 3D Ising
SG, than to $\beta$=1, predicted by the mean-field theory for
isotropic 3D Heisenberg SG. The quality of scaling is excellent,
as it can be seen in the bottom panel of Fig.~\ref{Scal17and50}.
The scaling for LSCNO50  is best for $\beta$=0.75, i.e. for the
same value as for LSCNO25. A comparable size of the $T$-region
where the scaling is valid starts for LSCNO50 at larger fields
than for LSCNO17 and LSCNO25, but the quality of the scaling is
still very good (see the upper panel in Fig.~\ref{Scal17and50}).

The parameter $\beta$, found here as a result of simple adjusting
procedure, is a critical exponent for the SG order parameter
$q_{EA}$, originally introduced by Edwards and Anderson in the
model based on the classical (mean-field)
calculations.\cite{Edwards75} The corresponding quantum-mechanical
calculations have been carried out by Fischer.\cite{Fischer75}
Since no lattice effects have been taken into account, the
obtained results describe the amorphous SG, where the CW law with
$\theta$=0 is found above $T_{SG}$.\cite{Fischer75} The
possibility that the average exchange interaction $J$ is not
zero, and thus causes that the ferromagnetic order competes with
the SG phase, has been included into the model by Sherrington,
Southern and Kirkpatrick (SSK), and the formula for extracting
$q_{EA}$ from the measured $\chi$ has been
given.\cite{Sherrington75,Sherrington75a,Kirkpatrick78} The
calculations based on the local-mean-field approximation suggest
that $q$ remains unchanged when $J$ changes its
sign.\cite{Patterson78} The one-component SSK model has been
modified and extended subsequently to describe a two-component
Ising-like magnetic system (with the separate order parameter for
each component), where the re-entrant transition to the SG phase,
both from the ferromagnetic and from the AF phase, is
predicted.\cite{Korenblit85,Takayama88,Fyodorov87} None of these
models provides a realistic description of LSCNO. Due to lack of
analytical expression for $q_{EA}$, the adjusting procedure seems
to be the best approach to find $\beta$.

\subsection{Dynamical scaling}
\label{sec:dynamical}
\begin{figure}
\includegraphics[width=0.50\textwidth, trim= 5 5 5 5]
{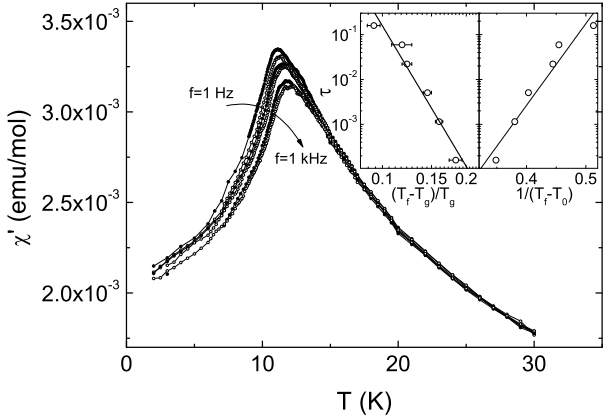}\caption{\label{acpeaks} The real part of
$\chi_{ac}$ as a function of temperature in the vicinity of the
transition to the SG phase at various frequencies (from top to
bottom: 1.00, 2.68, 7.19, 31.3, 138.7 and 997.3 Hz) for LSCNO25.
The bottom inset shows the peak positions for this sample (open
circles) together with those for LSCNO17 (solid squares)  and
LSCNO50 (solid diamonds). The thick solid lines are the best fits
to Eq.~\ref{power}. The details are described in the text and the
obtained parameters are given in Table \ref{table:slowing}. In the
upper inset, the slowing-down formula (Eq.~\ref{power}) versus
Vogel-Fulcher law (Eq.~\ref{VF}) is tested for description of
experimental data for LSCNO25. The thick (thin) solid line in left
(right) panel is the best fit to Eq.~\ref{power} (Eq.~\ref{VF}),
correspondingly. Note the logarithmic scale of abscissa in the
left panel and the linear one in the right panel.}
\end{figure}

A more detailed insight into the dynamics of
the SG state in LSCNO can be obtained from the analysis of the
$\chi_{ac}$ measurements. In Fig.\ref{acpeaks} we depicted the
temperature dependence of the real component of $\chi_{ac}$
for LSCNO25, measured with 1 Oe ac field amplitude
at various frequencies.
The data have been collected after cooling the sample
in zero field. The peak position $T_{f}$
moves toward higher temperature with increasing frequency. This
frequency dependence can be described by the standard critical
slowing down formula, given by the dynamic scaling
theory,\cite{Hohenberg77}
\begin{equation}
\tau=\tau_{0}\left(\frac{T_{f}-T_{g}}{T_{g}}\right)^{-z\nu}.\label{power}
\end{equation}
In this equation, the characteristic time $\tau$ describes the
dynamical fluctuation time scale and corresponds to the
observation time, $t_{obs}=1/\omega=1/2\pi f$, at the temperature
of maximum in $\chi'(T)$; $\tau_{0}$ is the shortest time
available to the system, i.e. the microscopic flipping time of the
fluctuating entities; $T_{f}$ is the frequency-dependent freezing
temperature and $\nu$ is the critical exponent of the spin (or
spin-cluster) correlation length $\xi$
($\xi\propto[T_{f}/(T_{f}-T_{g})]^{\nu}$). Below $T_{f}$, the
longest relaxation time of the system exceeds $t_{obs}$ and thus
the system is out of equilibrium. According to the dynamic scaling
hypothesis, the characteristic time $\tau$ in the vicinity of the
transition changes with the correlation length $\xi$ as
$\tau\propto\xi^{z}$.\cite{Ogielski85}

Fitting to the power law given by Eq.(\ref{power}) requires
adjusting three parameters: $\tau_{0}$, $T_{g}$ and the product
$z\nu$. Since $T_{g}$ is the infinitely slow cooling value of
$T_{f}$ (i.e. can be regarded as $\lim_{f \to 0} T_{f}$), we
carried out a more restrictive fit by assigning $T_{g}$ the value
of temperature at which the ZFC dc-$\chi(T)$ curve has its
maximum. Additionally, this allows us to overcome the difficulties
caused by the small number of the experimental points and the
relatively large errors compared with the small change of $T_{f}$
within 3 decades of the frequency variation. Having $T_{g}$ fixed,
it is possible to obtain directly $\tau_{0}$ and $z\nu$ from the
linear fits of log $\tau$ vs log $(T_{f}-T_{g})/T_{g}$. The best
fits, shown as the thick solid lines in the inset to
Fig.\ref{acpeaks} (left panel), yield the parameters given in
Table \ref{table:slowing}.

\begin{table}[ht]
\caption{The dynamic magnetic properties of LSCNO.}
\centering 
\begin{tabular*} {1.0\columnwidth}
{@{\extracolsep{\fill}}cccccc}
\hline\hline\\ 
& y=0.17&y=0.25&y=0.50\\  
[0ex]\\
\hline 
\\
$T_{g}$ (K) &6.8$\pm$0.1  & 10.4$\pm$0.1 & 13.1$\pm$0.1 \\ 
$\tau_{0}$ (s) & $10^{-9.5\pm0.4}$ & $10^{-10.5\pm1.0}$ & $10^{-9.7\pm0.5}$\\
$z\nu$ & -6.8$\pm$0.4 & -8.8$\pm$1.1 & -8.7$\pm$0.5\\
$E_{a}/k_{B}$ (K) & $\sim$7.6 & $\sim27$ & $\sim34$ \\
$T_{0}$ (K)  & 6.6$\pm$0.1  &  9.6$\pm0.2$ & 12.2$\pm0.1$\\
$\tau^{*}$ (s)  & $10^{-6.4\pm0.2}$ &  $10^{-8.3\pm0.4}$ & $10^{-7.8\pm0.3}$\\
[1ex] 
\hline\hline 
\end{tabular*}
\label{table:slowing} 
\end{table}

The typical values of $\tau_{0}$ for the canonical SG, i.e.
$10^{-12}-10^{-14}$ s, are of the order of the spin-flip time of
atomic magnetic moments ($\sim$$10^{-13}$
s).\cite{Wang2004,Vijayanandhini2009,Laiho2001,Gunnarsson88} As it
can be seen in Table \ref{table:slowing}, the LSCNO17 exhibits the
slowest dynamics among the investigated samples. Its
characteristic relaxation time is of the order of $\sim$$10^{-10}$
s and is evidently larger than $\sim$$10^{-13}$ s expected for the
single atomic spins. This strongly suggests the existence of spin
clusters. Even the shortest $\tau_{0}$ for LSCNO system, found for
$y$=0.25, is of the order of $\sim10^{-11}$ s and thus does not
exclude the existence of spin clusters, albeit the number of spins
in the fluctuating entities is expected to be
smaller.\cite{Mukadam2005,Mathieu2005}

The dynamic magnetic properties of a glassy system may be tested
in the frame of Vogel-Fulcher (VF) law,\cite{Tholence1980}
\begin{equation}
\tau=\tau^{*}\text{exp}[E_{a}/k_{B}(T_{f}-T_{0})].\label{VF}
\end{equation}
Taking the LSCNO25 data as an example, we show in the right panel
of the upper inset to Fig.\ref{acpeaks} that the change of the
relaxation time $\tau$ in LSCNO in the frequency range, which has
been experimentally accessible to us, can be described by this
formula equally well as by the power law given by
Eq.(\ref{power}).

The problem of discrimination between these two laws, the power
law given by Eq.(\ref{power}) and the VF law given by
Eq.(\ref{VF}), was a subject of debate two decades ago and
Eq.(\ref{power}) has been found to describe the experimental data
for the SG systems better than the VF law provided that the range
of $\tau$ is sufficiently large.\cite{Souletie85} However, the
difference is clearly visible only when the variation of $\tau$
approaches 11 orders of magnitude (see Fig.2 in
Ref.[\onlinecite{Souletie85}]). This probably explains why the
phenomenological VF law is still used in literature to describe
the frequency dependence of $T_{f}$ in the SG systems. The
high-$T$ expansion of the VF law is identical with that of the
power law up to terms of order $(T_{0}/T)^{3}$. Closer to $T_{0}$
(and $T_{g}$), the VF law can be adjusted to match a power law
over a large frequency range through the relation\cite{Souletie85}
\begin{equation}
ln\frac{40k_{B}T_{f}}{E_{a}}\sim\frac{25}{z\nu}.\label{adjustment}
\end{equation}

With $T_{f}$$\sim$11.6 K  found for LSCNO25 in the frequency range
of the experiment, the formula (\ref{adjustment}) gives
$E_{a}/k_{B}\sim27$ K for this system. Taking this value  of
$E_{a}$ as granted, we have fitted Eq.~(\ref{VF}) to the data. The
best fit, marked in the right panel of the upper inset to
Fig.~\ref{acpeaks} as the solid line, yields
$\tau^{*}=10^{-8.3\pm0.4}$ s and $T_{0}=9.6\pm0.2$ K. The
analogical analysis for LCNO17 and LSCNO25 leads to the parameters
given in Table \ref{table:slowing}. The values of $T_{0}$ are
smaller than $T_{g}$ from the power law (Eq.~\ref{power}) only by
a few percent, in accordance with the general trend found in the
metallic SG systems.\cite{Souletie85}

The phenomenological parameter $T_{0}$ has been introduced into
the standard Arrhenius law for glasses to overcome the
difficulties with keeping the magnitude of $\tau^{*}$ at a
physically-meaning level. However, $\tau_{0}$ from
Eq.~(\ref{power}) are believed to give more reliable insight into
the SG dynamics than $\tau^{*}$ does.\cite{Souletie85,Mukadam2005}

Despite this, the $T_{0}$ can be interpreted as a measure of the
coupling between the interacting entities.\cite{Shtrikman81} In
the frame of this picture, $T_{0}\ll E_{a}/k_{B}$ indicates a weak
coupling and $T_{0}\gg E_{a}/k_{B}$ a strong one. As it can be
seen in Table \ref{table:slowing}, $T_{0}$ for LSCNO varies from
$\sim$0.4~$E_{a}/k_{B}$ to $\sim$0.9~$E_{a}/k_{B}$. Thus $T_{0}$
is in the intermediate regime and confirms the conclusion drawn in
Sec.~\ref{ac} from the simple parameter $\delta$ about the
presence of some interactions between the magnetic entities in the
system, albeit does not settle whether they are single spins or
spin clusters. The energy scale $k_{B}T_{0}$ is almost equal to
$E_{a}$ for LSCNO17. This betokens the largest coupling between
the magnetic entities among the examined Ni concentrations. This
is consistent with the observation that $|\theta|$ in the CW law,
describing $\chi(T)$ in the paramagnetic region, achieves its
maximal value in the LSCNO system for $y$=0.15-0.17 (the inset to
Fig.~\ref{CW}) and suggests that the local AF order, partially
restored by trapping the mobile holes, is strongest at this Ni
concentration.

\subsection{Role of Ni in the Cu-O network}
\label{sec:Ni} As it is widely recognized, a 2D dynamic AF order
persists even in the overdoped LSCO,\cite{Birgeneau88} so the
dynamical AF fluctuations are present in LSCO15. They may be
regarded as a reminiscence of a 3D static AF order in the parent
compound La$_{2}$CuO$_{4}$, destroyed by the frustrating impact of
the mobile-hole spins introduced by Sr doping.\cite{Aharony88}
Near the famous $x$=1/8 anomaly, the neutron scattering
experiments reveal a gap in the low-energy spin excitation
spectrum, and presence of the magnetic incommensurate peaks that
has been interpreted as an evidence for formation of the static
stripes of spins and holes in the Cu-O planes.\cite{Tranquada95,
Kofu2009} In LSCO15, this static stripe order disappears but the
spin gap of 4 meV is still visible in the neutron measurements.
Introduction of even a small number of Ni atoms ($y$$<$0.03) into
the system reduces the spin-excitation energy
scale.\cite{Kofu2005}

Our study clearly shows that the low-temperature phase of Ni-doped
LSCO15 exhibits all thermodynamical characteristics of the SG
phase, including the static scaling. The dynamical scaling
parameters suggest the presence of the ordered magnetic clusters.
The XAFS measurements reveal that the Ni ions added to the Cu-O
planes act as the hole absorbers.\cite{Hiraka2009} Such Ni$^{2+}$
ion with the trapped hole has been suggested to form the
Zhang-Rice doublet\cite{Zhang88} with the effective spin $S$=1/2
and not to disturb the AF correlations between the $S$=1/2 spins
of the Cu$^{2+}$ ions.\cite{Hiraka2009} Our observation of the SG
behavior suggests that the compensation of the $S$=1/2 Cu$^{2+}$
spin vacancy in the Cu-O network by the effective spin of the
complex formed by the Ni$^{2+}$ ion and the bounded ligand hole is
not perfect. The observed localized magnetic moment for
$y$$<$0.09, attributed to the Ni site, is tiny (corresponding to
$S$=0.11) but finite. The $T_{g}$ extrapolated into the SC region
of the phase diagram takes finite values (see the inset to
Fig.~\ref{ZFCandFC}), what implies that the frustrating effects
appear in the system with the first added Ni ions. Thus, on the
contrary to the conclusions of Ref.~[\onlinecite{Hiraka2009}], our
study shows that Ni exhibits a magnetic character and affects the
magnetic correlations between Cu spins in the Cu-O planes starting
from its smallest concentrations.

The picture of a glassy system depends on the time scale of the
used measuring probe. No static component in the magnetic signals
has been revealed by neutron experiments on LSCNO up to $y$=0.029.
On the other hand, formation of the short-range magnetic order in
the vicinity of the Ni ions is strongly suggested even in the
lightly doped LSCO15 by the zero-field $\mu$SR experiments
revealing slowing down of the Cu spins
fluctuations.\cite{Adachi2008} For larger Ni content, the
procession of muon spins, indicating the long-range magnetic
order, has been clearly observed for $y$=0.10 at 0.3 K and the
static magnetic order has been suggested to form even for
$y$$\geq$0.07 at temperatures below 2 K, outside our measurement
window.\cite{Adachi2008} It should be remembered that any
relaxation slower than 10$^{-5}$-10$^{-6}$ s is seen as a "static"
component by $\mu$SR.\cite{Binder86,Adachi2004} Moreover, the
damping rate in the $\mu$SR time-spectra term representing the
muon-spin procession has been found to increase with increasing Ni
content $y$ for $y$$\geq$0.03 in the Ni-doped LSCO with $x$=0.13
(LSCO13). Based on this, the less coherent magnetic order, seen by
$\mu$SR experiments in LSCO13 with $y$=0.10, has been suggested to
be a precursory state toward the SG state.\cite{Adachi2004} Our
direct observation of the SG behavior in Ni-doped LSCO15 is in
agreement with this conjecture and our results are consistent with
the $\mu$SR measurements.

As presented in Sec.~\ref{dc}, the susceptibility of LSCNO with
large $y$ exhibits the CW dependence with the negative $\theta$.
It is striking that the $|\theta|$ reaches its maximal value at
around $y$=0.15-0.17, i.e. where the Ni concentration is equal to
the hole concentration in the system (see the inset to
Fig.~\ref{CW}). This can be qualitatively understood as a result
of trapping mobile holes by Ni.

In the framework of this model, trapping the mobile holes in
LSCO15 by Ni ions restores \textit{locally} the AF order in the
Cu-O planes.\cite{Tsutsui2009,Hiraka2009} The more Ni ions are
present in the system, the more mobile holes become localized and
the restored AF order is stronger. This is reflected as the
increase of absolute value of the effective $\theta$ with $y$.
When all mobile holes became localized, i.e. $y$ becomes equal to
the hole concentration, there is no way to increase the AF
correlations between the Cu spins and this explains why $|\theta|$
does not continue to increase with increasing $y$ above
$y$=0.15-0.17.

As mentioned in Sec.~\ref{dc}, the increase of the calculated
$\mu_{eff}$ per Ni ion with increasing $y$ for $y$$>$0.07 means
that some Ni ions are not in the Ni$^{2+}$L hole-bounded state at
these concentrations. Thus, a small number of the Ni$^{2+}$ ions
in $S$=1 state is probably present in the system even for
$y$$<$0.15. However, their influence is not able to overwhelm the
effect of restoring locally the AF correlations between the Cu
spins. When the number of Ni ions is sufficiently large to
localize all mobile holes, adding the further Ni atoms to the
system means introducing the subsequent $S$=1 magnetic moments
into the existing Cu-O network with the locally AF ordered
regions. This destroys the restored local AF correlations between
the Cu spins what is observed as decrease of $|\theta|$ with
increasing $y$ above 0.15-0.17.

The SG transition temperature, $T_{g}$, in LSCNO increases
linearly with increasing $y$ up to $y$=0.30, as shown in the inset
to Fig.~\ref{CW}. In the cluster SG phase of LSCO, the monotonic
decrease of $T_{g}$ with increasing hole concentration $x$ for
0.03$\leq$$x$$\leq$0.05 has been tried to explain by the
finite-size-scaling hypothesis.\cite{Cho92,Niedermayer98}
According to this proposition, $T_{g}$ is expected to be
proportional to the size of the locally ordered regions $L^{d}$
($d$=2), determined by the concentration $x$ of the  mobile holes
destroying the AF order, $T_{g}\propto L^{2}\propto1/x$, in rough
agreement with the experimental
results.\cite{Niedermayer98,Julien03}

Applying this model directly to LSCNO would mean assuming that the
average size of the magnetically ordered region increases with
increasing $y$ even up to the limit when 1/3 of the Cu ions is
replaced by the Ni ions. However, decrease of $|\theta|$ with
increasing $y$ above $y$$\simeq$0.17, presumably reflecting the
destroying impact of the $S$=1 Ni ions on the locally restored AF
order, suggests that this is not true. Thus, the finite-size
effects are probably not the only factors determining the value of
$T_{g}$ in LSCNO up to $y$$\approx$0.30.

At large $y$, magnetism of LSCNO is presumably dominated by the Ni
ions with the spin $S$=1. As it can be seen in the inset to
Fig.~\ref{CW}, the $T_{g}$ continue to increase with increasing
$y$ above $y$$\approx$0.30, although with the smaller rate than
for $y$$<$0.30, and exhibits a quasi-linear dependence on $y$ up
to the largest examined Ni concentration. The situation bears some
analogy to the standard metallic magnetic alloys, where the linear
dependence $T_{g}$ on the impurity concentration $y_{im}$ is found
at large $y_{im}$ (in addition to the similar
$T_{g}$$\propto$$y_{im}$ dependence in the regime of the
interacting single spins at low $y_{im}$$<$0.5\%, followed by the
$T_{g}$$\propto$$y_{im}^{2/3}$ relationship at larger
$y$).\cite{Mydosh93} This takes place in the interval from about
10 at.\% to the magnetic percolation limit (where a smooth
transition to the inhomogeneous long-range order appears). The
magnetism of the glassy system with $y_{im}$ within this
concentration interval is dominated by the large ordered clusters
embedded in the nonmagnetic SG matrix.\cite{Mydosh93} In LSCNO,
the $S$=1 Ni ions are embedded in the very complex magnetic
background but probability of forming clusters by these ions and
the average size of such clusters increases with increasing $y$.
Thus, the magnetic behavior of LSCNO might be dominated by these
clusters at large $y$. The observed increase of the characteristic
time for the fluctuating entities in the system, $\tau_{0}$, with
increasing $y$ from 0.25 to 0.50 (see Table \ref{table:slowing})
is consistent with such supposition.

Alternatively, one may interpret the observed SG behavior in LSCNO
in the framework of the stripe pinning picture. Following this
concept, it was suggested that the dynamical stripe correlations
of spins and holes are localized in the vicinity of Ni ions in
LSCO15 leading to the formation of the static stripe
order.\cite{Adachi2004,Adachi2008}

However, the results of the remnant magnetization $M_{\text{TRM}}$
measurements, presented in detail in Sec.~\ref{remnant}, give us
some grounds for speculations contrary to the above
interpretation. The $M_{\text{TRM}}$ decay in LSCNO is described
by a stretched exponential function with $1-n\approx1/3$ exponent,
in perfect agreement with the theoretical and experimental results
for the canonical SG
systems.\cite{Campbell86,Campbell88,Ogielski85,Chamberlin84} In
"pure" LSNO (without Cu), where the presence of stripes has been
unambiguously confirmed experimentally, the time dependence of the
"isothermal" remanent magnetization (i.e. obtained after ZFC,
applying field at 2 K and next switching the field off again), is
inconsistent with a stretched exponential
function.\cite{Freeman2006} On this ground, it was concluded that
the mechanism by which the LSNO come back to equilibrium is not
the same as in the canonical SG, at least in the measured LSNO
with $x$=1/3 Sr content where the stripes are commensurate with
the square lattice of the Ni-O planes, i.e. where the charge and
magnetic orders have the identical periods.\cite{Freeman2006}

At the opposite side of the LSCO-LSNO phase diagram, in the SG
phase of "pure" (without Ni) LSCO with $x$=0.04, the same time
dependence of $M_{\text{TRM}}$ as that found in LSCNO, and with
the same value of $1-n\approx1/3$, has been reported.\cite{Chou95}
However, it has been underscored that the duration of the
experiment had not been sufficient to exclude the different types
of $M_{TRM}(t)$ dependence.\cite{Chou95} In LSCNO, the
sufficiently long time of the measurement and the standard
analysis allow us to exclude those other possible forms of the
$M_{\text{TRM}}$ vs $t$ dependence (see Sec.~\ref{remnant}).

To recapitulate, the above results can suggest that the origin of
the SG phase in LSCNO is not related to the (potential) presence
of the stripes and disorder in their array because the mechanism
of relaxation seems to be identical with the one observed in the
canonical SG and different from that observed in the reference
stripe-ordered compound LSNO.

\subsection{Critical exponents}
\label{exponents} In general, the critical exponents are believed
to be universal and thus capable of describing  the behavior of the
system near phase transition even when the detailed microscopic
picture is not known. As regards exponent $z\nu$ from the dynamical
scaling [Eq.~\ref{power}], the numerical simulations for the 3D Ising SG
model give $z\nu=7.9$ (while for a conventional phase transition
$z\nu=2$ is expected).\cite{Ogielski85} Experimentally, the values
of $z\nu$ between 5 and 11 have been observed for different SG
systems\cite{Fischer91_ch8,Souletie85} and thus $z\nu\sim7-9$
found for LSCNO (see Table \ref{table:slowing}) are inside the SG
realm and close to the value predicted for the Ising-like systems.
However, it should be remember that it is not possible to
distinguish significantly different $z\nu$ values on systems of
different spin dimensionality.\cite{Nordblad2004} For instance,
$z\nu\sim10-11$ has been reported both for a short-range Ising SG
system,
Fe$_{0.5}$Mn$_{0.5}$TiO$_{3}$,\cite{Gunnarsson88,Mattsson95} as
well as for more isotropic vector SG: namely, for a 3D XY system,
Eu$_{0.5}$Sr$_{1.5}$MnO$_{4}$,\cite{Mathieu2005} and for 3D
Heisenberg-like AuFe$_{8\%}$.\cite{Souletie85,Lundgren82}

With fitted $\delta$ and adjusted $\beta$, together with the
dynamical critical exponent $z\nu$, the other critical exponents
can be calculated using the scaling equations, in the similar
manner to $\gamma$ from Eq.~(\ref{delta}). Namely, the heat-capacity
exponent $\alpha$ can be calculated now from the equation
\begin{equation}
\alpha+2\beta+\gamma=2 \label{alpha},
\end{equation}
and the spin-correlation length exponent $\nu$ can be obtained
from the relationship
\begin{equation}
d\nu=2-\alpha,\label{nu}
\end{equation}
where $d$ is the dimensionality of the magnetic interactions.
Since the measurements of the canonical SG/metal multilayers have
showed that even a very weak magnetic coupling between layers of
2D SG causes a 3D character of the whole system,\cite{Djurberg95}
we have taken $d$=3 in this equation for LSCNO. The known value of
$\nu$ allows to extract the correlation-time exponent, $z$, from
the results of the dynamical slowing-down formula fitting. The
exponent $\eta$ that governs the spatial correlation function at
$T_{g}$ can be deduced from the equation
\begin{equation}
(2-\eta)\nu=\gamma.\label{eta}
\end{equation}

The whole set of the obtained critical exponents, including these
from the dynamical scaling, is given in Table
\ref{table:exponents}.
\begin{table}[ht]
\caption{The critical exponents for LSCNO together with the values
reported for 3D Heisenberg SG (Ref.[\onlinecite{Levy88}]) and
these predicted by numerical simulations for 3D Ising system
(first row for a given exponent - Ref.[\onlinecite{Ogielski85}],
second row -
Refs.[\onlinecite{Ballesteros2000}],[\onlinecite{Campbell2006}]
  and [\onlinecite{Hasenbusch2008}]).}
\centering 
\begin{tabular*} {1.0\columnwidth}
{@{\extracolsep{\fill}}cccccc}
\hline\hline\\ 
&Heisenberg &y=0.17&y=0.25&y=0.50&Ising\\  
[0ex]\\
\hline 
\\
$\delta$ & 3.3  & 6.8$\pm$0.1 & 5.8$\pm$0.1 & 6.0$\pm$0.1 & 6.8\\ 
& &  & & &8--9 \\
$\beta$ & 0.9 & 0.55$\pm$0.05 & 0.75$\pm$0.05 & 0.75$\pm$0.05 &0.5 \\
&  & & & &0.7--0.8\\
$\gamma$ & 2.3 & 3.2$\pm$0.4 & 3.6$\pm$0.3 & 3.8$\pm$0.4 & 2.9\\
& & & & &5.0--6.5\\
$\alpha$ & -2.1  & -2.3$\pm$0.5 & -3.1$\pm$0.4 & -3.3$\pm$0.5 &-1.9 \\
& & & & &-4.5-- -6.1\\
$\nu$  & 1.3  &  1.4$\pm0.2$ & 1.7$\pm0.1$ & 1.8$\pm0.2$ & 1.3\\
  & & & & & 2.2--2.7\footnote[1]{Refs.[\onlinecite{Ballesteros2000}],[\onlinecite{Campbell2006}]
  and [\onlinecite{Hasenbusch2008}]. The rest of static exponents corresponding to these values of
  $\nu$ and $\eta$ are obtained
  using scaling and hyperscaling relations.} \\
$\eta$  & 0.4 & -0.23$\pm$0.01 & -0.12$\pm$0.02&
 -0.14$\pm$0.02 & -0.22 \\
& & & & &  -0.34--~-0.40\footnotemark \\
$z\nu$  & 7  & 6.8$\pm0.4$ & 8.8$\pm$1.1 & 8.7$\pm$0.5 & 7.9\\
$z$  & 5.4  & 4.8$\pm$1.1 & 5.2$\pm1.0$ & 5.0$\pm$0.7 & 6.1 \\
[1ex] 
\hline\hline 
\end{tabular*}
\label{table:exponents} 
\end{table}
The corresponding values for the 3D Heisenberg and 3D Ising
systems are listed for comparison. The critical exponents for AgMn
from Ref. [\onlinecite{Levy88}], given in the first column of
Table \ref{table:exponents}, are very typical for the canonical
SG, being weakly anisotropic Heisenberg-like SG, and thus can
serve as a reference point for the Heisenberg-class
systems.\cite{Campbell2010,Kawamura2010,Mathieu2005}. Since a good
experimental realization of the Ising SG seems to be
difficult,\cite{Kawamura2010} we have used the results of the
numerical simulations made by Ogielski,\cite{Ogielski85} revised
by some later large-scale
computations,\cite{Ballesteros2000,Campbell2006,Hasenbusch2008} as
a reference point for this universality class. The measured values
for FeMnTiO$_{3}$, regarded as the best laboratory realization
of the Ising system so far, are roughly consistent with these numerical
predictions.\cite{Gunnarsson91}

A large variation of the critical-exponent values measured for the
same SG exists in the literature, partially because of the
different field and temperature ranges used.\cite{Fischer91} Thus
any attempt to classify a given system to one of the universality
classes, based solely on these experimental values, should be
taken always with some caution. In the single crystals of "pure"
cuprate La$_{1.96}$Sr$_{0.04}$CuO$_{4}$, $\delta$=5.9$\pm$0.6 and
$\gamma$=4.3$\pm$1.4, similar to the values for LSCNO, have been
reported.\cite{Chou95} They are significantly larger than those
observed in the Heisenberg-like SG (see Table
\ref{table:exponents}) and close to the exponents measured in the
Ising-like systems: 2D Rb$_{2}$Cu$_{1-x}$Co$_{x}$F$_{4}$
($\gamma$=4.5$\pm$0.2) and 3D Fe$_{0.5}$Mn$_{0.5}$TiO$_{3}$
($\gamma$=4.0$\pm$0.3, $\delta$=8.4$\pm$0.15). However, the
$\chi(T)$ curve in La$_{1.96}$Sr$_{0.04}$CuO$_{4}$ bifurcates
below $T_{g}$ both for \textit{H}$\parallel$\textit{ab}-plane and
for \textit{H}$\parallel$\textit{c} and thus the system has been
classified as a 3D Heisenberg-like one.\cite{Chou95}

Essentially, all exponents lie between those for Ising-like and
Heisenberg-like SG. The values for $y$=0.25 are practically the
same as for $y$=0.50 and no obvious trend is visible with changing
$y$. In the single crystals of "pure" nickelate
La$_{2-x}$Sr$_{x}$NiO$_{4+\delta}$ ($0.02\leq x \leq0.5$) the
difference between the FC and ZFC magnetization curves has been
found only for
\textit{B}$\parallel$\textit{ab}-plane.\cite{Freeman2006} Taking
into account the aforementioned different result for the "pure"
cuprate La$_{1.96}$Sr$_{0.04}$CuO$_{4}$, one might naively expect
an evolution of the LSCNO properties from the isotropic to more
anisotropic ones with increasing $y$. No such clear evolution is
reflected in the values of exponents for LSCNO.

Let us note that the critical exponents for LSCNO are very similar
to those reported for Eu$_{0.5}$Sr$_{1.5}$MnO$_{4}$, classified as
a 3D XY SG because the bifurcation of $\chi(T)$ curve and
frequency dependence of the peak position in ac susceptibility has
been found only for the field in the $ab$ plane.\cite{Mathieu2005}
This suggests that LSCNO may belong to the XY SG universality
class. However, without measurements on single crystals or at
least on the magnetically ordered powders no definite conclusion
can be made.

\section{Conclusions}
\label{sec:conclusions} The magnetic nature of a Ni dopant in LSCO
cannot be neglected, even when its concentration is small. The
low-temperature phase of LSCNO exhibits all the features that
characterize the spin-glass behavior. The $\chi(T)$ curve displays
a ZFC-FC bifurcation and a distinct peak is seen in the ZFC
branch. A stretched-exponential decay of the thermoremnant
magnetization takes place below the transition temperature. The
position of a peak in the real part of the ac-susceptibility,
which is accompanied by a step in the imaginary component, is
frequency-dependent and this dependence is described by the
standard critical slowing-down formula. The characteristic time
that governs the internal dynamics of the system indicates
possible existence of clusters. The nonlinear part of the dc
susceptibility exhibits scaling behavior characteristic for
the SG. All these features - when taken together - confirm
presence of the SG state in LSCNO at low temperatures. The
critical exponents lie between those characteristic for
Ising-like and Heisenberg-like systems.

The transition temperature decreases linearly with decreasing Ni
content $y$ and extrapolates to 0 K at $y$=0, suggesting that Ni
exhibits a magnetic character and affects the AF correlations
between Cu spins in the Cu-O layers starting from the smallest
concentration $y$, in the superconducting region of LSCNO phase
diagram. It is possible that the (cluster) SG phase coexists with
the local AF order partially restored by trapping holes at Ni
sites.

\begin{acknowledgments}
This work was supported by funds of the Polish Ministry of Science
and Higher Education as a research project for years 2007-2011
(Grant No N202 048 32/1183). We would like to thank Weigang Wang
for the part of ac susceptibility measurements.
\end{acknowledgments}

\begin{appendix}
\section{Apical oxygen atoms}
\label{sec:apical} 
\begin{figure}
\includegraphics[width=0.50\textwidth, trim= 5 5 5 5]
{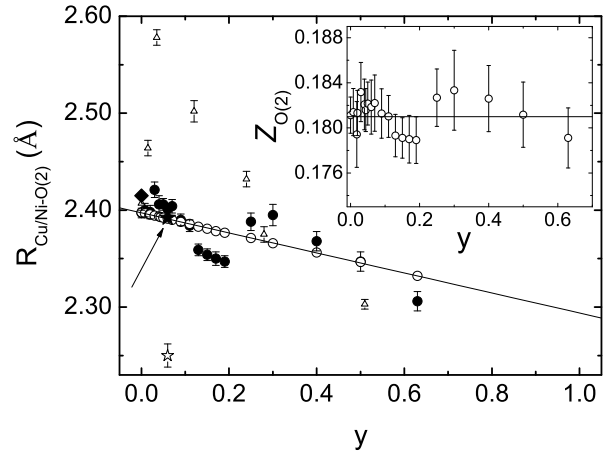}\caption{\label{apical} The average distance
between Cu/Ni and apical oxygen atom in LSCNO with various Ni
content, $y$. The solid circles mark the direct results of
Rietveld refinement procedure while the open circles denote the
same results with the assumption that $z_{\text{O(2)}}=0.181$~\AA=
const. The solid line is the linear interpolation of the results
marked as open circles. The small open triangles mark the data
from Ref.~[\onlinecite{Wu96}], the large solid diamond - the value
of $R_{\text{Cu-O(2)}}$ from Ref.~[\onlinecite{Radaelli94}], and
the open star - $R_{\text{Ni-O(2)}}($y$=0.06)$ from
Ref.~[\onlinecite{Haskel2001}]. The arrow indicates
$R_{\text{Cu/Ni-O(2)}}($y$=0.06)$, marked as the solid star and
calculated as described in the text. Inset: the refined $z$
coordinate of O(2) apical atom. The solid line marks the value
0.181.}
\end{figure}

The Rietveld refinement procedure allows to
determine the values of free atomic coordinates. The position of
the apical oxygen atom O(2) in the tetragonal unit cell of LSCNO
is (0,0,$z$). The $z_\text{{O(2)}}$ values as a function of Ni
content are shown in the inset to Fig.~\ref{apical}.
The X-ray diffraction measurements are not light-atoms sensitive;
therefore the uncertainties of $z_\text{{O(2)}}$ are relative
large. Within the error margins, there are no obvious trend in the
$z_\text{{O(2)}}$ vs $y$ dependence and $z_\text{{O(2)}}$ can be
accepted as being constant and equal to 0.181(3) in the whole
investigated doping range. The uncertainties in the determined
$z_\text{{O(2)}}$ are transferred onto the calculated
\textit{average} interatomic Cu/Ni-O(2) distances,
$R_\text{{Cu/Ni-O(2)}}\equiv R_{\text{O(2)}}$. Thus, the found
$R_{\text{O(2)}}$ vs $y$ behavior mimics $z_\text{{O(2)}}$ vs $y$
dependence, as it can be clearly seen in the main panel of
Fig.~\ref{apical}. However, taking $z_\text{{O(2)}}$=const as
granted results in smooth linear dependence of $R_{\text{O(2)}}$
vs $y$ (see open circles in Fig.~\ref{apical}) that mimics linear
decrease of $c$ with increasing $y$. This means that within the
error margins our measurements do not reveal a nonmonotonic
$R_{\text{O(2)}}$ vs $y$ behavior observed in
Ref.~[\onlinecite{Wu96}]. As mentioned in Sec.~\ref{sec:crystal},
XPD is not atom-specific
in the sense that only \textit{averaged} Cu/Ni-O(2) distance is
measured. For a given $y$, this measured $R_\text{{O(2)}}$
distance can be expressed as $y\cdot R_\text{{Ni-O(2)}}+(1-y)\cdot
R_\text{{Cu-O(2)}}$, where $R_\text{{Ni-O(2)}}$
($R_\text{{Cu-O(2)}}$) is Ni-O(2) [Cu-O(2)] distance,
respectively. $R_{\text{O(2)}}$ for $y$=0.06 can be calculated in
this way with the use of $R_\text{{Ni-O(2)}}$=2.250(12)
\AA~obtained for this Ni content by Haskel \textit{et al.} from
EXAFS spectra (Ref.~[\onlinecite{Haskel2001}]) and
$R_\text{{Cu-O(2)}}$=2.397(5) \AA~~from our XPD measurements for
undoped ($y$=0) structure. The calculated
$R_{\text{O(2)}}$($y$=0.06)=2.392 \AA~~is in perfect agreement
with the value 2.391 \AA~~obtained directly from Rietveld analysis
with the assumption that $z_{\text{O(2)}}=0.181$~\AA~~(see
Fig.~\ref{apical}). No significant differences were found in the
$R_\text{{Ni-O(2)}}$ values for
$y\leq0.07$.\cite{Haskel2001,Hiraka2009} The linear dependence of
$R_\text{{Cu/Ni-O(2)}}$ vs $y$, when extrapolated to $y$=1, yields
the value that is 0.04 \AA~~larger than
$R_\text{{Ni-O(2)}}$($y$=0.06) (see the solid line in
Fig.~\ref{apical}). This fact gives some grounds for speculation
that $R_\text{{Ni-O(2)}}$ may change for $y>0.07$.

\section{Scaling analysis}
\label{AnotherScaling} Since the different approaches to the
static scaling in SG exist in
literature,\cite{Dekker88,Chou95,Geschwind90} it is worth to
explore them one after the other to check, which of them describes
the experimental data in the best way. In the following, we will
present the details of scaling analysis for LSCNO25, but we have
also carried out the similar analysis for LSCNO17 and LSCNO50.

In the derivation of the Eq.~\ref{scal1} the interactions in the
system are assumed to average perfectly to
zero.\cite{Barbara81,Fischer75,Suzuki77} Since evidently this is
not our case and $\theta$ in the CW law describing $\chi_{l}$ of
LSCNO has a substantial value, we first tried to follow the
approach based on expanding magnetization $M$ in odd powers of
$\chi_{l}H$ (instead of
$H/T$)\cite{Omari83,Yeung87,Dekker88,Gingras96}
\begin{equation}
M(T,H)=\chi_{l}H-a_{3}(\chi_{l}H)^{3}+a_{5}(\chi_{l}H)^{5}
-a_{7}(\chi_{l}H)^{7}+...\label{expanding},
\end{equation}
The coefficients $a_{3}$, $a_{5}$,... here are $T$-dependent.
Since now $\chi_{l}$ has not a simple Curie form,
$\chi_{l}\propto1/T$, the scaling described by Eq.~(\ref{scal1})
is replaced by the following relationship\cite{Dekker88}
\begin{equation}
\frac{\chi_{nl}}{\mid
t\mid^{\beta}}=g_{\mp}\left(\frac{\chi_{l}H}{|t|^{\frac{\beta+\gamma}{2}}}\right)
\label{scal2},
\end{equation}
where $g_{-}$ and $g_{+}$  is another pair of the scaling
functions.

\begin{figure}
\includegraphics[width=0.50\textwidth, trim= 0 0 0 0]
{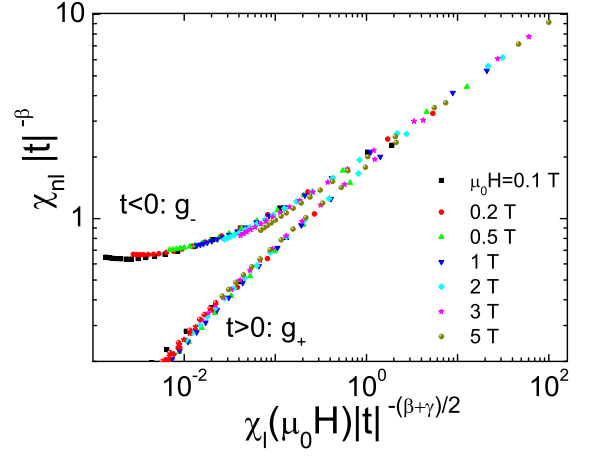}\caption{\label{scal2Fig} (Color online)
Scaling plot for LSCNO25 according to Eq.~(\ref{scal2}) with
$\beta$=0.75 and $\gamma$=3.6 for the nonlinear susceptibility
$\chi_{nl}$ at various magnetic fields from 0.1 T to 5 T at
temperatures below ($t$$<$0) and above ($t$$>$0) $T_{g}$=10.2 K.}
\end{figure}

Since both pair of scaling functions, $f_{\mp}(x)$ from
Eq.~(\ref{scal1}) and $g_{\mp}(x)$ from Eq.~(\ref{scal2}), behave
as const$\cdot x^{2/\delta}$ in the large-$x$ limit,
\cite{Chalupa77,Barbara81,Dekker88} the same value
$\delta$=5.8$\pm$0.1 found from the power dependence of
$\chi_{nl}$ vs $H$ at $T_{g}$ (Eq.~(\ref{delta}); see also the
inset to Fig.~\ref{Fields}) is used in both scalings to calculate
$\gamma$ from the adjusted $\beta$ [Eq.~(\ref{ScalLaw})]. To
perform scaling according to Eq.~(\ref{scal2}), $\beta$ is
adjusted in such a way that coincidence of the data on two curves
in $\chi_{nl}/|t|^{\beta}$ vs $\chi_{l}H/|t|^{(\beta+\gamma)/2}$
plot is as good qualitatively as possible. The best results have
been obtained for $\beta$=0.75 and $\gamma$=3.6. As it is seen in
Fig.\ref{scal2Fig}, the quality of the scaling is not
satisfactory.
\begin{figure}[H]
\includegraphics[width=0.50\textwidth, trim= 0 0 0 0]
{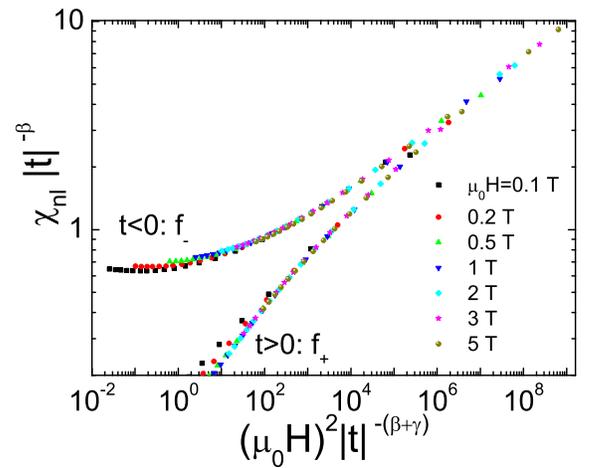}\caption{\label{scal1Fig} (Color online)
Scaling plot for the nonlinear susceptibility $\chi_{nl}$ of
LSCNO25 according to Eq.~(\ref{scal1}) with $\beta$=0.75 and
$\gamma$=3.6 and at the same fields and temperatures as in
Fig.~\ref{scal2Fig}.}
\end{figure}

To improve the quality of scaling we have employed
Eq.~(\ref{scal1}). The best collapse of the data onto the two
universal curves has been obtained once again for $\beta$=0.75
with the uncertainty estimated to be $\sim$0.05. This implies
$\gamma$=3.6$\pm$0.3. The resulting scaling is shown in
Fig.\ref{scal1Fig}. The quality of scaling is much better than
this obtained by employing Eq.~(\ref{scal2}) and almost perfect
for fields larger than 0.2 T. Thus, this form of scaling [i.e.
given by Eq.~(\ref{scal1})] is used for LSCNO.

However, while the log-log plots presented in Figs.~\ref{scal2Fig}
and \ref{scal1Fig} are enough to make direct comparison between
the quality of obtained the best scalings according to
Eq.~(\ref{scal2}) or Eq.~(\ref{scal1}), to estimate the critical
temperature region where the scaling is valid, it is better to use
Eq.~(\ref{scal1a}), in which the argument of the scaling function
is linear in $t$, as described in Sec.~\ref{sec:static}.

\end{appendix}

\end{document}